\documentclass[iop, apj, twocolappendix, numberedappendix, appendixfloats]{emulateapj}
\usepackage{threeparttable}
\usepackage{multirow}
\usepackage{times}
\usepackage{enumitem}
\usepackage{url} 
\usepackage{amsmath,bm}

\usepackage{aas_macros}

\usepackage{hyperref}
\hypersetup{
    colorlinks=true,
    linkcolor=blue,
    filecolor=magenta,      
    urlcolor=blue,
    pdftitle={Overleaf Example},
    pdfpagemode=FullScreen,
    }

\shorttitle{}
\shortauthors{GU H. et al.}
\begin{document}

\title{The Stellar Abundances and Galactic Evolution Survey (SAGES). II. Machine Learning-Based Stellar parameters for 21 million stars from the First Data Release}

\author{\href{https://orcid.org/0009-0007-5610-6495}{Hongrui Gu}\altaffilmark{1,2}}
\author{\href{https://orcid.org/0000-0002-6790-2397}{Zhou Fan}$^{\ast}$\altaffilmark{1,2}}
\author{\href{https://orcid.org/0000-0002-8980-945X}{Gang Zhao} \altaffilmark{1,2}}
\author{\href{https://orcid.org/0000-0003-3250-2876}{Yang Huang}\altaffilmark{1,2}}
\author{\href{https://orcid.org/0000-0003-4573-6233}{Timothy C. Beers}\altaffilmark{3,4}}
\author{\href{https://orcid.org/0000-0002-9702-4441}{Wei Wang}\altaffilmark{1}}
\author{\href{https://orcid.org/0000-0001-6637-6973}{Jie Zheng}\altaffilmark{1}}
\author{\href{https://orcid.org/0000-0003-2868-8276}{Jingkun Zhao}\altaffilmark{1}}
\author{Chun Li\altaffilmark{1}}
\author{\href{https://orcid.org/0000-0002-8442-901X}{Yuqin Chen}\altaffilmark{1}}
\author{\href{https://orcid.org/0000-0003-2471-2363}{Haibo Yuan}\altaffilmark{5}}
\author{\href{https://orcid.org/0000-0002-0389-9264}{Haining Li}\altaffilmark{1}}
\author{\href{https://orcid.org/0000-0003-0173-6397}{Kefeng Tan}\altaffilmark{1}}
\author{\href{https://orcid.org/0000-0001-7255-5003}{Yihan Song}\altaffilmark{1}}
\author{\href{https://orcid.org/0000-0001-7865-2648}{Ali Luo}\altaffilmark{1}}
\author{Nan Song\altaffilmark{6}}
\author{Yujuan Liu\altaffilmark{1}}

\altaffiltext{1}{CAS Key Laboratory of Optical Astronomy, National Astronomical Observatories, Chinese Academy of Sciences, Beijing 100101, P.R.China; {\it zfan@nao.cas.cn}}%; {\it gzhao@nao.cas.cn}
\altaffiltext{2}{School of Astronomy and Space Science, University of Chinese Academy of Sciences, Beijing, P.R.China}
\altaffiltext{3}{Department of Physics and Astronomy, University of Notre Dame, Notre Dame, IN 46556, USA}
\altaffiltext{4}{Joint Institute for Nuclear Astrophysics -- Center for the Evolution of the Elements (JINA-CEE), USA}
\altaffiltext{5}{Department of Astronomy, Beijing Normal University, Beijing 100875, People's Republic of China;}
\altaffiltext{6}{China Science and Technology Museum, Beijing, 100101, People's Republic of China;}

%\footnote{Corresponding auther: Gang Zhao. 
\begin{abstract}
Stellar parameters for large samples of stars play a crucial role in constraining the nature of stars and stellar populations in the Galaxy. An increasing number of medium-band photometric surveys are presently used in estimating stellar parameters. In this study, we present a machine-learning approach to derive estimates of stellar parameters, including [Fe/H], log $g$, and $\it T$$\rm _{eff}$, based on a combination of medium-band and broad-band photometric observations. Our analysis employs data primarily sourced from the SAGE Survey , which aims to observe much of the Northern Hemisphere.
We combine the $uv$-band data from SAGES DR1 with photometric and astrometric data from Gaia EDR3, and apply the random forest method to estimate stellar parameters for approximately 21 million stars. We are able to obtain precisions of 0.09\,dex for [Fe/H], 0.12\,dex for log $g$, and 70\,K for $\it T$$\rm _{eff}$. Furthermore, by incorporating 2MASS and WISE infrared photometric and GALEX ultraviolet data, we are able to achieve even higher precision estimates for over 2.2 million stars. These results are applicable to both giant and dwarf stars. Building upon this mapping, we construct a foundational dataset for research on metal-poor stars, the structure of the Milky Way, and beyond. With the forthcoming release of additional bands from SAGE Survey such DDO51 and H-$\alpha$, this versatile machine learning approach is poised to play an important role in upcoming surveys featuring expanded filter sets.
\end{abstract}
\section{Introduction}\label{sec1}
Accurate determination of stellar parameters for large samples of stars in our galaxy plays an important role in many fields, including stellar physics and the structure and evolution of the Milky Way. Spectroscopic observations can accurately determine stellar parameters from spectral lines. As early as 2009, the Sloan Extension for Galactic Understanding and Exploration (SEGUE; \citealt{2009AJ....137.4377Y}) project, part of the Sloan Digital Sky Survey (SDSS; \citealt{2000AJ....120.1579Y}), used low-resolution ($R \sim 1800$) spectra to obtain stellar parameters for some 240,000 stars. Results for an additional 118,958 unique stars were reported on by \citet{2022ApJS..259...60R}. Subsequently, the SDSS APOGEE project (\citealt{2017AJ....154...94M}) obtained stellar parameters for over 700,000 stars using high-resolution ($R \sim 22,500$) near-infrared spectra. With the advent of the Large Sky AreaMulti-Object Fiber Spectroscopy Telescope (LAMOST, \citealt{2012RAA....12..723Z},\citealt{2012RAA....12.1197C}), multi-fiber spectroscopic survey telescopes have greatly improved the efficiency of acquiring spectra. To date, LAMOST has observed more than 11 million stars, with over 7 million stars with available parameters, increasing the sample size by an order of magnitude. However, compared to photometric survey projects with over a billion stars at limiting magnitudes of around $G = 20$, such as Gaia DR3 \citep{2021A&A...649A...1G}, there is still a one to two order of magnitude gap between the sample sizes of photometric surveys and spectroscopic surveys, underscoring the importance of obtaining stellar-parameter estimates through the use of photometry.

Stellar-parameter estimates based on multi-band photometric surveys have expanded over the past several decades. The HM (\citealt{1998A&AS..129..431H}) and the GCS (\citealt{2004PASA...21..129N}) surveys, contemporaneous with the Sloan Digital Sky Survey, included medium-band filters, but had limiting magnitudes of only $v \sim$ 8, far from meeting current research needs.  \cite{2008ApJ...684..287I} used spectral and photometric data from SDSS to estimate metallicities for some 200,000 FGK dwarfs, with a precision of about 0.2-0.3\,dex down to [Fe/H] $\sim -2.0$. With improvements in photometric accuracy and precision, \cite{2015ApJ...803...13Y} estimated metallicities for some 500,000 FGK dwarfs to a level of 0.1-0.2\,dex in SDSS/Stripe 82 down to [Fe/H] $\sim -2$;  \cite{2021RAA....21..319Z} demonstrated that estimates for giant stars can also achieve 0.2\,dex precision down to a similar limit, with differing model parameters compared to FGK stars. However, estimating parameters of giant stars using dwarf-star models could introduce a certain degree of systematic bias (\citealt{2020ApJ...897...39A}).
Space-based telescopes offer higher photometric accuracy and more uniform survey coverage. \cite{2022ApJS..263...29X} used Gaia EDR3 data to provide metallicity estimates for 27 million FGK stars across the sky with 0.2\,dex precision. 
Due to the bandwidth limitations of wide-band photometric survey filters, it is difficult to significantly improve upon metallicity and other stellar-parameter estimates further using such filters alone. 

The Skymapper Southern Sky Survey (SMSS, \citealt{2007ASPC..364..177K}), led by the Australian National University, has adopted the SC filter system (\citealt{1964ApNr....9..333S}, \citealt{1970AJ.....75..624C}), similar to the GCS and HM surveys in some of its filters, covering as faint as 20th magnitude (at 5 $\sigma$) in the Southern Hemisphere over 21,000 square degrees. \cite{2019ApJS..243....7H} used polynomial fitting to establish empirical relationships between stellar parameters and photometric colors, deriving accurate atmospheric parameters for about 1 million red giants from SMSS DR1.1. \cite{2021ApJS..254...31C} used a grid-based synthetic-photometry method to obtain over 250,000 photometric metallicities for giants from SMSS DR2. Subsequently, using re-calibrated SMSS DR2 data and Gaia EDR3 data, \cite{2022ApJ...925..164H} employed polynomial-fitting methods to produce a catalog of stellar parameters, including metallicity, for about 24 million stars with a precision of 
0.1-0.3\,dex, with carefully selected training datasets enabling metallicity estimates down to [Fe/H] $\sim$ -3.5. 

By installing a single narrow-band filter with a response curve centered around the Ca II H and K lines on the CFHT telescope, the Pristine Survey (\citealt{2017MNRAS.471.2587S}) is capable of observing thousands of square degrees to a depth of about 20th magnitude, while maintaining a 0.1-0.2 \,dex precision in metallicity estimates down to [Fe/H] $\sim -3.5$ in 
their DR1 (\citealt{2023arXiv230801344M}).

With the advent of multiple narrow-band photometric surveys, such as J-PLUS (\citealt{2019A&A...622A.176C}), S-PLUS (\citealt{2019MNRAS.489..241M}), and J-PAS (\citealt{2021A&A...653A..31B}), traditional grid-based, parameter-fitting, and other empirically based algorithms become increasingly difficult to handle in such high-dimensional scenarios. Therefore, various machine-learning algorithms have been employed in the measurement of stellar parameters. Examples include neural networks (\citealt{2020MNRAS.499.5447K}, \citealt{2021ApJ...912..147W}, \citealt{2022A&A...659A.181Y}), and random forest algorithms (\citealt{2019AJ....158...93B}, \citealt{2022A&A...657A..35G}). These studies have demonstrated that machine-learning algorithms are highly effective methods for extracting stellar parameters from high-dimensional data.

The Stellar Abundance and Galaxy Evolution Survey (SAGES; \citealt{2014IAUS..298..326W}, \citealt{2018RAA....18..147Z}, \citealt{2019RAA....19....3Z}, \citealt{2022RAA....22j5004T}, \citealt{2023ApJS..268....9F}) is a photometric survey in the Northern Hemisphere that includes narrow- and medium-band filters. Its $v_s$-band central wavelength coincides with the location of the  Ca II H and K lines, and its 
$u$-band is similar to SMSS, providing the opportunity to obtain more accurate atmospheric parameters, and reducing the impact of molecular carbon bands. 
In addition to the $u$- and $v$-band data used in this work, broad-band photometric data, such as SDSS-like $g$, $r$, $i$, and medium-band data, such as DDO51 and H-$\alpha$, will continue to be released. In this approach, an algorithm must be chosen that can be applied to both low-dimensional data and complex high-dimensional data. This is necessary, since the current techniques (e.g., \citealt{2023ApJ...957...65H}) of measuring stellar parameters based on DR1 data cannot be iterated with the release of additional filter bands in the SAGES survey. Robust statistical methods have played a significant role in handling high-dimensional astronomical datasets, and it is logical to expect that they would perform similarly well or even better in low dimensions. By applying such statistical methods to low-dimensional datasets, such as multi-band photometric data from SAGES, it is possible to simultaneously attain high-precision stellar parameters, and gain experience for the future exploration of large multi-dimensional photometric datasets.

This paper is organized as follows. Section \ref{sec2} describes the data and methods used in this work. Section \ref{sec3} introduces the results for the training data. Section \ref{sec4} describes the stellar parameters derived for SAGES stars, followed by a summary in Section \ref{sec5}.

\begin{figure}
\begin{center}
\includegraphics[scale=0.4,angle=0]{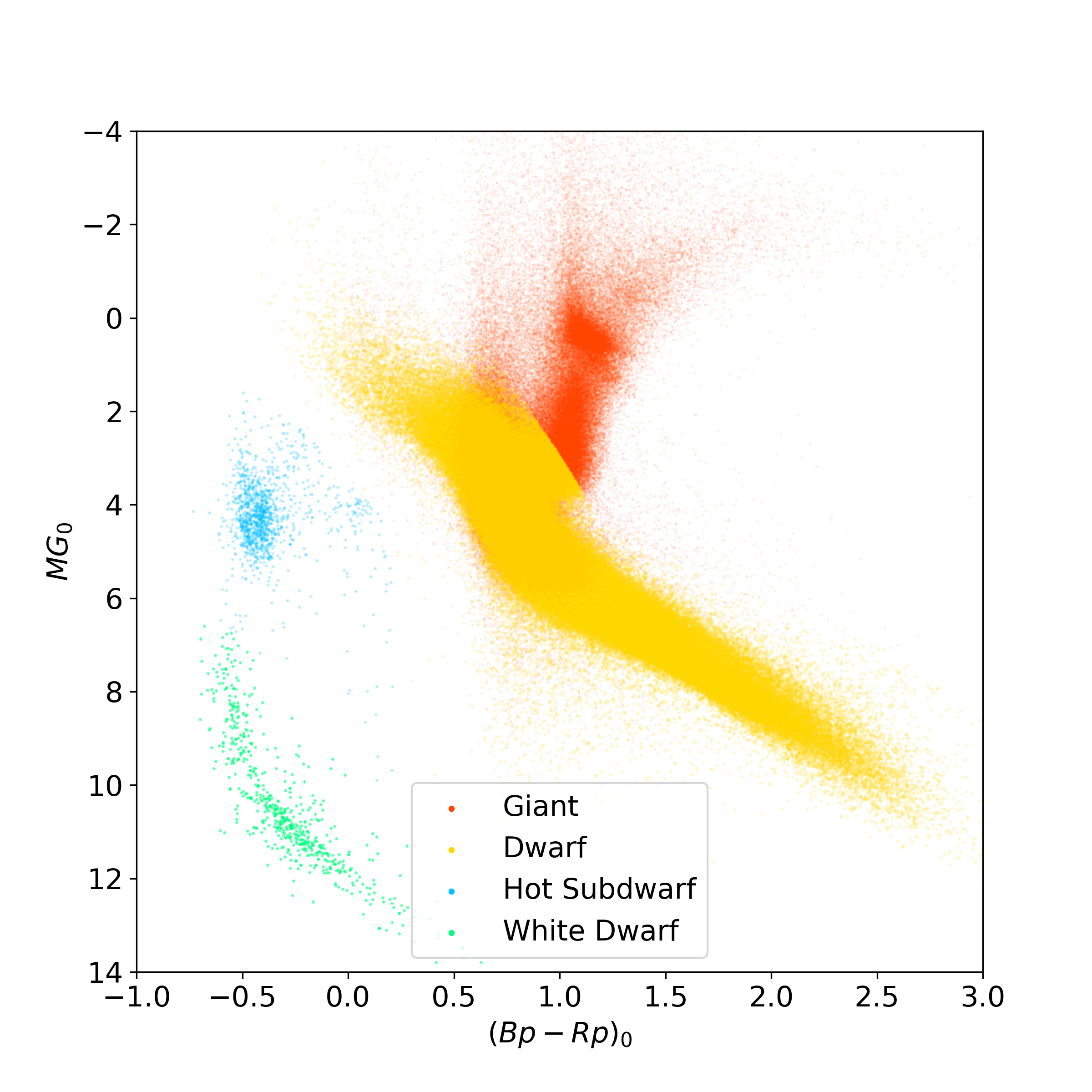}
\caption{Distribution of the data used in this work in the Gaia color-absolute magnitude diagram. The colors distinguish four different types of stars. The data used in this work are the dwarf (yellow) and giant (red) stars.} \label{Fig1}
\end{center}
\end{figure}

\section{Data and methods}\label{sec2}

In order to train the random forest algorithm, we constructed two sets of training data, including the colors and filter magnitudes provided by SAGES and Gaia and the estimated stellar parameters provided by spectroscopic surveys, as described below. 

%The photometric data is used as the input of the algorithm, and the output is compared with the stellar parameters to finish the training.

\subsection{Photometric Data}

The first training set uses photometric data from SAGES DR1 and Gaia EDR3 (\citealt{2021A&A...649A...1G}) as input; the second set additionally includes the photometric data in the infrared from 2MASS (\citealt{2003tmc..book.....C}) and ALLWISE (\citealt{2014yCat.2328....0C}) and ultraviolet from GALEX (\citealt{2014yCat.2335....0B}). SAGES is a multi-band photometric survey focused on estimation of stellar parameters and interstellar extinction. Its narrow $v$-band filter is quite sensitive to [Fe/H]. Its $u$-band and $v$-band are located on either side of the Balmer jump, providing good sensitivity to surface gravity (hereafter, log $g$). Gaia EDR3 provides a large amount of accurate photometric information (the ultra-wide $Bp$- and $Rp$-bands) for the stars in SAGES; so it can provide effective temperature estimates (hereafter, $T_{\rm eff}$). In addition, infrared photometric information such as ALLWISE and 2MASS (\citealt{2014ApJ...797...13S}), ultraviolet photometric information such as GALEX (\citealt{2024ApJS..271...26L}) can also be used to identify candidate metal-poor stars, taking advantage of the sensitivity of the infrared and ultraviolet bands to metallicity . 

In this work, two sets of training data are constructed.  The first set of data are chosen to apply to as large a sample of SAGES stars as possible, while providing parameter estimates with reasonable accuracy and precision. The second set of data sacrifices sample size in order to improve the accuracy and precision as much as 
possible for a subset of the SAGES stars. 

For the extinction in the various photometric bands, we employ the reddening coefficients provided by \cite{2022RAA....22j5004T} to correct the SAGES $u$- and $v$-band data, and the coefficients provided in Table 2 of \cite{2023ApJS..264...14Z} to correct the rest of data. All 
values of reddening coefficient we used are shown in Table 1. The extinction map we used is SFD98 (\citealt{1998ApJ...500..525S}).

Figure \ref{Fig1} is the color-magnitude diagram (x: ($Bp-Rp$)$_0$, y: $M_{G0}$) for SAGES stars, after application of the reddening and extinction corrections.  We have divided the targets into dwarf stars (yellow dots), giant stars (red dots), hot subdwarfs (blue dots), and white dwarfs (green dots). Targets with Gaia parallax errors larger than the parallax itself are considered to be likely giants. Table 2 lists the constraints we imposed on the input training data and the output prediction data. The total sample size of the first set of data, where we emphasize retaining as large a sample as feasible, is 21,071,305, including 19,663,040 dwarfs and 1,408,265 giants. The total sample size of the second set of data, emphasizing accuracy and precision of the derived stellar parameters, is 2,191,452, including 2,037,528 dwarf stars and 153,924 giants.

\begin{table}
\centering
\caption{Reddening Coefficients for Photometry Data Used in this Work}
\begin{tabular}{llcl}
\hline
\hline
Band&Survey&Reddening&Citation\\
&&coefficient&\\
\hline
$NUV$&GALEX&7.294&\cite{2023ApJS..264...14Z}\\
$u$&SAGES&4.324&\cite{2022RAA....22j5004T}\\
$v$&SAGES&3.917&\cite{2022RAA....22j5004T}\\
$G$&Gaia&2.364&\cite{2023ApJS..264...14Z}\\
$Bp$&Gaia&2.998&\cite{2023ApJS..264...14Z}\\
$Rp$&Gaia&1.737&\cite{2023ApJS..264...14Z}\\
$J$&2MASS&0.748&\cite{2023ApJS..264...14Z}\\
$H$&2MASS&0.453&\cite{2023ApJS..264...14Z}\\
$K$&2MASS&0.306&\cite{2023ApJS..264...14Z}\\
$W$1&WISE&0.194&\cite{2023ApJS..264...14Z}\\
$W2$&WISE&0.138&\cite{2023ApJS..264...14Z}\\
\hline
\end{tabular}
\end{table}

\subsection{Stellar Parameters from Spectroscopic Surveys}

Spectroscopic observation is the most direct way to obtain stellar parameters. Even relatively low-resolution spectra can be used to estimate stellar parameters with higher accuracy and precision than current photometric methods, in particular if they are of high signal-to-noise. LAMOST (\citealt{2012RAA....12.1197C}),
as the largest spectroscopic survey project in the Northern Hemisphere at present, can provide the three stellar parameters required in this work: $T_{\rm eff}$, log $g$, and [Fe/H]. However, through DR10, LAMOST spectra were only low-resolution ($R \sim 1800$). The accuracy and precision of the stellar parameters in DR10 is lower than that of the medium- or high-resolution spectra, and the most metal-poor star reliably estimated by most pipelines in DR10 
%(WHICH ONES, CONTRAST WITH BETTER PIPLEINE RESULTS ?) 
%The standard pipeline of LAMOST employs template matching to determine metallicities, with their templates currently reaching as low as -2.5\,dex for the most metal-poor stars.
is only be about [Fe/H] $\sim -2.5$. In order to obtain training stars with lower metal abundance, we considered using samples from multiple spectral surveys. We employ samples from other 3 spectroscopic sky surveys: 
(1) APOGEE DR17 \citep{2023yCat.3286....0A}, a high-resolution near-IR survey with wavelengths ranging from 1.51 to 1.70 $\mu$m and $R \sim 22,500$, 
(2) RAVE DR5 \citep{2017AJ....153...75K}, a spectroscopic survey aiming to measure radial velocity with wavelengths centered on the Ca I triplet (8410\,{\AA} to 8795\,{\AA}) and resolving power of $R \sim 7500$, 
(3) PASTEL \citep{2016A&A...591A.118S}, a catalog of collected stellar parameters from the literature obtained by various high-resolution spectroscopic observations. 
(4) A number of additional papers have reported on results based on high-resolution follow-up observations for very metal-poor (VMP; [Fe/H] $\leq -2.0$) and extremely metal-poor (EMP; [Fe/H] $\leq -3.0$) stars identified in previous surveys that are not included in the PASTEL catalog.  We collected a total of 15,628 stars from \cite{2018ApJS..238...16L}, \cite{2004AAS...205.5210L}, \cite{2009A&A...501..519B}, \cite{2013ApJ...762...26Y}, \cite{2013AJ....145...13A}, \cite{2016ApJ...833...20Y}, \cite{2018A&A...620A.187F},  \cite{2020ApJ...898..150E}, and \cite{2021ApJ...907...10L}. These stars have metallicities in the range [$-0.1$, $-7.8$]; 93.1$\%$ are VMP stars and 9.8$\%$ are EMP stars.

\begin{table}
\centering
\caption{Adopted Constraints on the Datasets for Training and Prediction}
\begin{tabular}{lcc}
\hline
\hline
Parameter&Training cut&Prediction cut\\
\hline
Parallex over error&1&1\\
E(B-V)&0.05&0.5\\
$T_{\rm eff}$$_{err}$ &200& \dots \\
log $g_{err}$ &0.3& \dots \\

[Fe/H]$_{err}$ &0.2& \dots \\
$u_{err}$&0.1&0.2\\
$v_{err}$&0.1&0.2\\
$G_{err}$&0.01&0.1\\
$Bp_{err}$&0.01&0.1\\
$Rp_{err}$&0.01&0.1\\
$J_{err}$&0.04&0.1\\
$H_{err}$&0.05&0.1\\
$K_{err}$&0.05&0.2\\
$W1_{err}$&0.03&0.1\\
$W2_{err}$&0.05&0.1\\
$NUV_{err}$&0.1&0.2\\
\hline
\end{tabular}
\end{table}

\begin{figure*}
\begin{center}
\includegraphics[scale=0.33,angle=0]{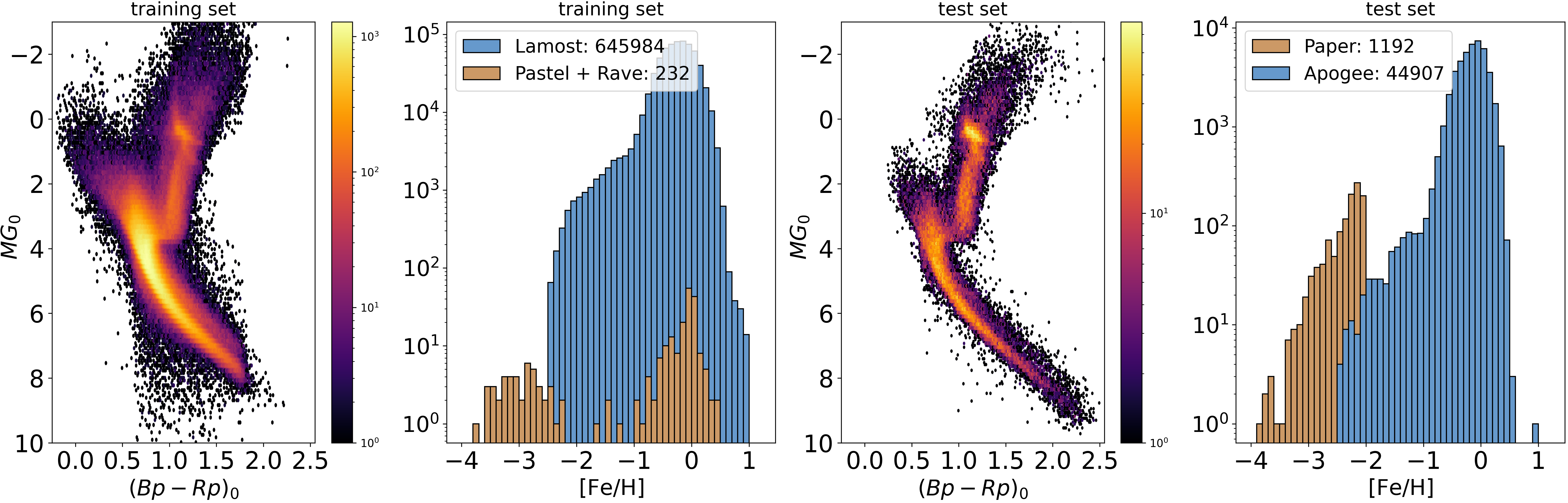}
\caption{{\it \textbf{Left two panels:}} Number-density distribution of the training samples for Data Set 1 in the color-absolute magnitude diagram, and the histogram of [Fe/H] from different spectral survey catalogs.
{\it \textbf{Right two panels:}} Same to left two panels but from test set.}
\end{center}
\end{figure*}
\begin{figure*}
\begin{center}
\includegraphics[scale=0.33,angle=0]{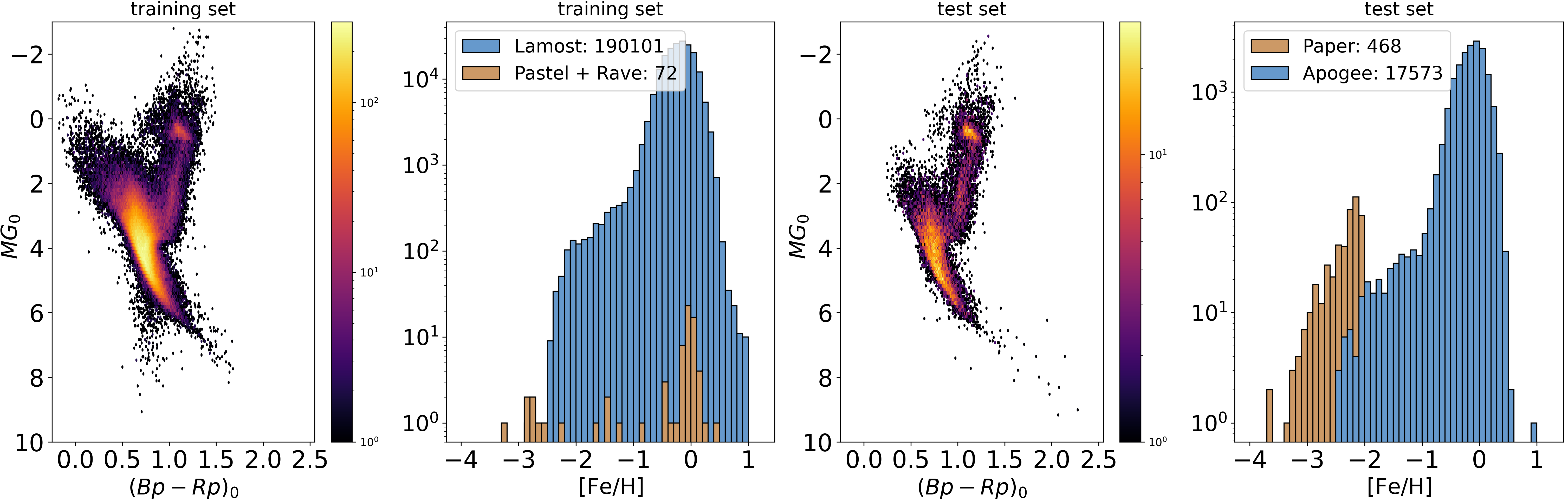}
\caption{Similar to Figure 2, but using data for Data  Set 2.}
\end{center}
\end{figure*}

\subsection{Training Samples}

To obtain stellar parameters from the SAGES photometric data, it is necessary to cross-match the stars in common between both spectroscopic surveys and SAGES (constraints are as shown in Table 2). During the training step, the apparent magnitudes, reddening corrections, and other parameters are used as input data; the stellar parameters provided by spectral data are used to constrain output predictions. To ensure the independence and the coverage of different parameters in the parameter space, this work uses LAMOST data as the training set and APOGEE data as the test set. To extend the parameter range at low-metallicity, data from Pastel and Rave are utilized. Additionally, data from other 9 papers are used to test the result at low-metallicity. Considering the systematic differences in the parameters provided by different datasets, a correction of 0.427 in [Fe/H] is applied to the data from 9 papers to account for these systematic differences. Figures 2 and 3 show the number density distribution of these two samples in the 
color-magnitude diagram and the number-density distribution of metallicity in training data and testing data.

\begin{figure*}
\begin{center}
\includegraphics[scale=0.5,angle=0]{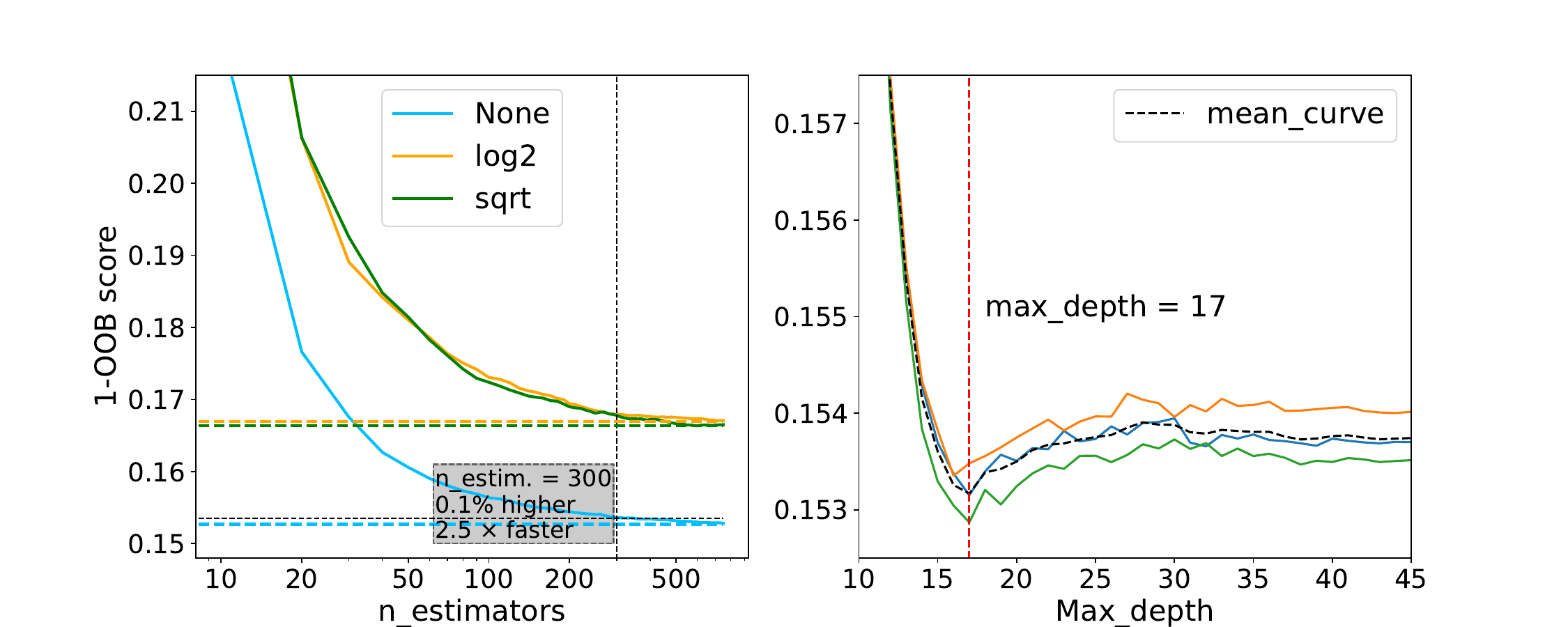}
\caption{{\it \textbf{Left panel:}} The Out-of-Bag (OOB) scores for different combinations of hyperparameters are shown, with different colors representing different max\_features values, and the x-axis representing different n\_estimators values. The y-axis represents 1 $-$ OOB score, so a smaller value on the y-axis indicates better performance. For the selected n\_estimators $=$ 300, the accuracy is only 0.1\% lower than when n\_estimators $=$ 750, but the computational speed is 2.5 times faster.
{\it \textbf{Right panel:}} For n\_estimators = 300 and max\_features $=$ 'None', the 1$-$OOB score varies with the max\_ depth parameter. A smaller value on the y-axis indicates better performance. By testing each max\_depth three times and averaging the results to obtain the mean curve (represented by the black dashed line), it can be seen that the optimal performance is achieved when max\_depth$=$17.}
\end{center}
\end{figure*}

\subsection{Method: Random Forest}
After compared the results of a simple three-layer fully connected neural network under different loss functions, polynomial fitting, and random forests method, we find out that the random forest algorithm not only achieved the fastest speed but also the best performance, slightly outperforming polynomial fitting and significantly outperforming the simple fully connected neural network. Therefore, this work chose to use the random forest algorithm (\citealt{2001MachL..45....5B}) for subsequent work.
Random forest is an ensemble-learning approach, assembled from multiple independent decision-tree algorithms. Each decision tree divides the space through a series of discriminants that seek to minimize the "entropy", defined as the Gini index. The decision trees are trained through sets of partial training samples; the final result is obtained by weighting the results of the multiple decision trees. The random forest has the advantage that is can prevent over-fitting, and thus is more generally applicable to a variety of input data.

Moreover, since each sample is drawn with replacement from the dataset to serve as the training set for each tree, the samples that are not selected can be used as a validation set, known as the out-of-bag (OOB) error. Therefore, when using a random forest, there is no need to establish an additional validation set to perform real-time evaluation of the model's performance.

Random forest, as a machine learning algorithm, has several hyperparameters, and selecting an appropriate set of these hyperparameters is crucial. In this work, two very important hyperparameters were chosen: n\_estimators and max\_features. n\_estimators refers to the number of decision trees in the random forest, while max\_features is the number of features considered in each decision tree. Ultimately, we selected n\_estimators = 300 and max\_features = 'None', which means all input features are used. After determining and fixed the most important two parameters, another hyperparameter, max\_depth, which represents the maximum depth of each individual decision tree, was tested. After testing different depths, the final value for max\_depth was determined to be 17. Details of above testing are shown in figure 4.

\subsection{Model Establishment}

Since the prediction accuracy and precision of a single model to simultaneously predict multiple parameters is generally less than that obtained by individual models to predict single parameters, we constructed multiple random forest models to predict the [Fe/H], $\it T$$\rm _{eff}$, and log $g$ for dwarf and giant stars, respectively. Surface gravity is one of the primary criteria to distinguish dwarf and giant stars; if we were to predict the log $g$ of dwarfs and giants stars separately, it would introduce some mis-classification error for the model. Therefore, we employ a unified model to predict the log $g$ of dwarf and giant stars together. Five random-forest models are trained for the two data sets in this work, which are models for: (1) [Fe/H] of dwarf stars, (2) [Fe/H] of giant stars, (3) $\it T$$\rm _{eff}$ of dwarf stars, (4) $\it T$$\rm _{eff}$ of giant stars, and (5) log $g$ for both dwarfs and giants.  The  flow chart is shown in Figure 5.

\begin{figure*}
\begin{center}
\includegraphics[scale=0.3,angle=0]{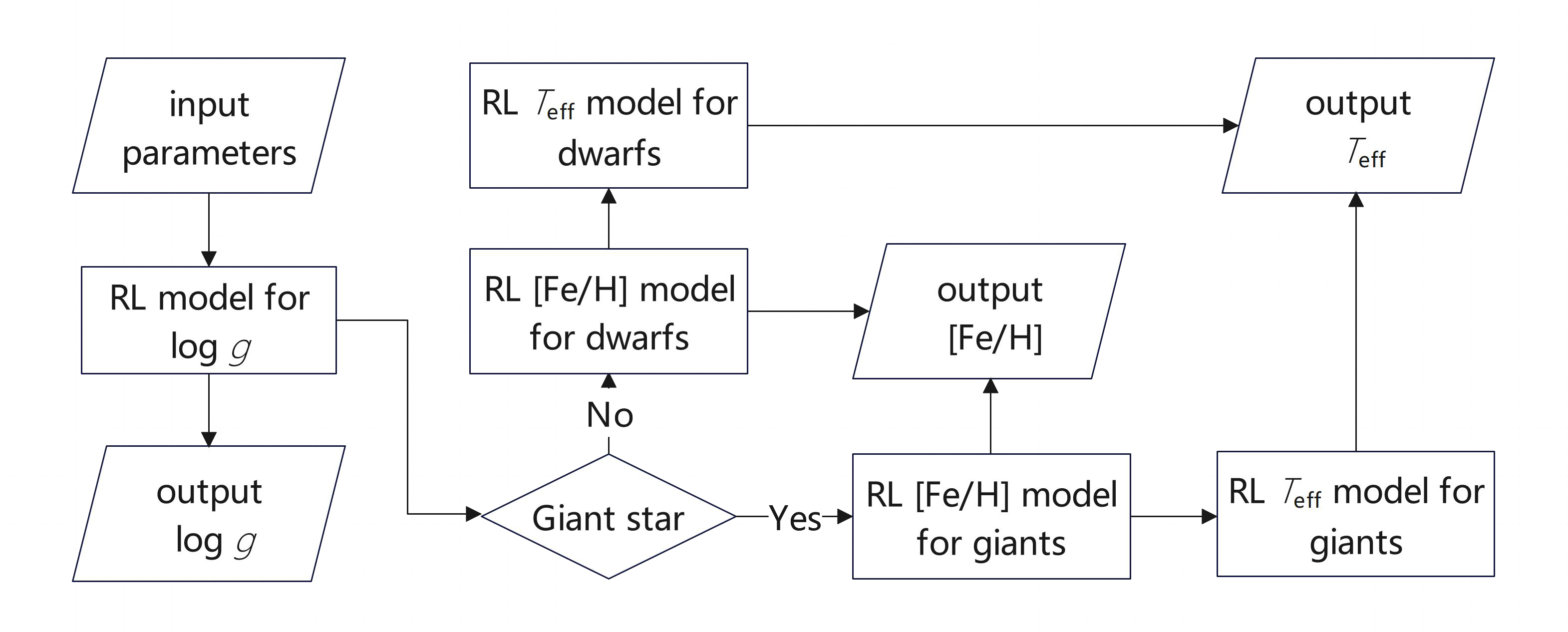}
\caption{Flow chart of this work.}
\end{center}
\end{figure*}

The input data for these models include the magnitude after extinction correction, the colors after reddening correction, the parallex, the $G$-band absolute magnitude, and $E(B-V)$. The magnitudes and $E(B-V)$ can provide the model with extinction, reddening, and effective temperature, and the $G$-band absolute magnitude and colors can provide the model with information similar to the color-magnitude diagram, which not only constrain the model$\it T$$\rm _{eff}$ and stellar-classification information, but also provides constraints on the derived log $g$. The $u$- and $v$-band photometry and related colors mainly provide information for the models to constrain the estimated [Fe/H]. Because the metallic absorption lines cover both the visible, ultraviolet and near-infrared spectra, visible photometry from Gaia (\citealt{2022ApJS..263...29X}), ultraviolet photometry from GALEX (\citealt{2024ApJS..271...26L}), and infrared photometry from WISE and 2MASS (\citealt{2014ApJ...797...13S}) can also provide information on the metallicity of stars. 
During the training, we used parameters from LAMOST as the training set and APOGEE as the test set. Additionally, we supplemented the training set with metallicity data from PASTEL and RAVE for metal-poor stars, and we used the metallicity data from nine papers to supplement the metal-poor stars in the test set. Due to the systematic difference between nine papers and the LAMOST, this work applied a systematic correction of 0.427 to the metallicity values from nine papers.

\section{Results}\label{sec3}

We now compare the results for the two sets of training samples. The first data set includes the photometric information provided by SAGES DR1 and Gaia EDR3, and the second data set uses information from SAGES DR1, Gaia EDR3, WISE, 2MASS, and GALEX.

\begin{figure*}
\begin{center}
\includegraphics[scale=0.50,angle=0]{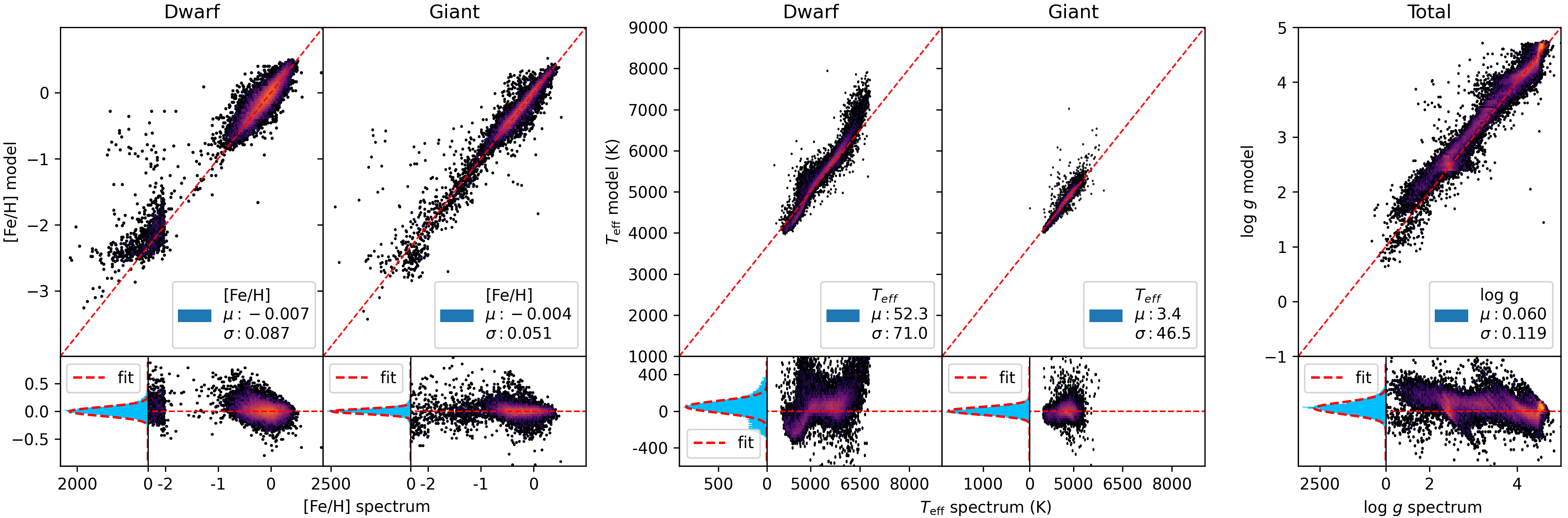}
\caption{Comparisons between the spectroscopic and photometric stellar parameters for the test set from Data Set 1, for 5 models (dwarf [Fe/H], giant [Fe/H], dwarf $\it T$$\rm _{eff}$, giant $\it T$$\rm _{eff}$, log $g$). The lower portion of each panel shows the number density distribution of the difference between two spectroscopic and photometric stellar parameters. The mean and standard deviation are obtained from fitting the histograms, and their values are shown in the legends.
}
\end{center}
\end{figure*}

%\vskip 1cm
\subsection{Data Set 1: Photometry from SAGES DR1 and Gaia DR3}

The training data for this input data contains 579,185 dwarf stars and 67,031 giant stars. Figure 6 shows the accuracy and precision of the five models on the test set; each subplot represents the training result for one model. The upper part of each panel compares the known stellar parameters derived from spectroscopy (x-axis) with the parameters predicted by the model (y-axis). The left portion of the lower part in each panel represents the Gaussian fitting results for the residuals. The right portion of the lower part in each panel shows the residual distribution of the predicted values minus the spectroscopic values, as a function of the spectroscopic values. The fitting parameters are shown in the legend in the upper part of each panel. 

%The titles of the small plot and the stellar parameters shown in the legend indicate the model used in this plot. 

\begin{figure*}
\begin{center}
\includegraphics[scale=0.50,angle=0]{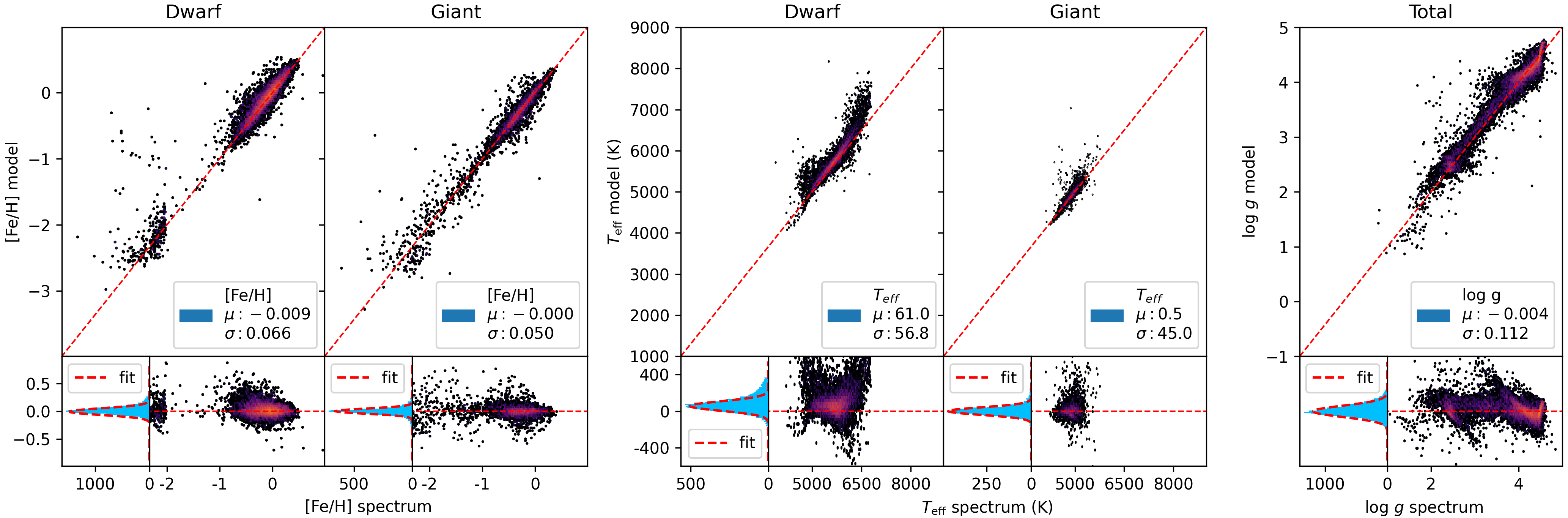}
\caption{Similar to Figure 6, but using the test set from Data Set 2.}
\end{center}
\end{figure*}
\begin{figure*}
\begin{center}
\includegraphics[scale=0.42,angle=0]{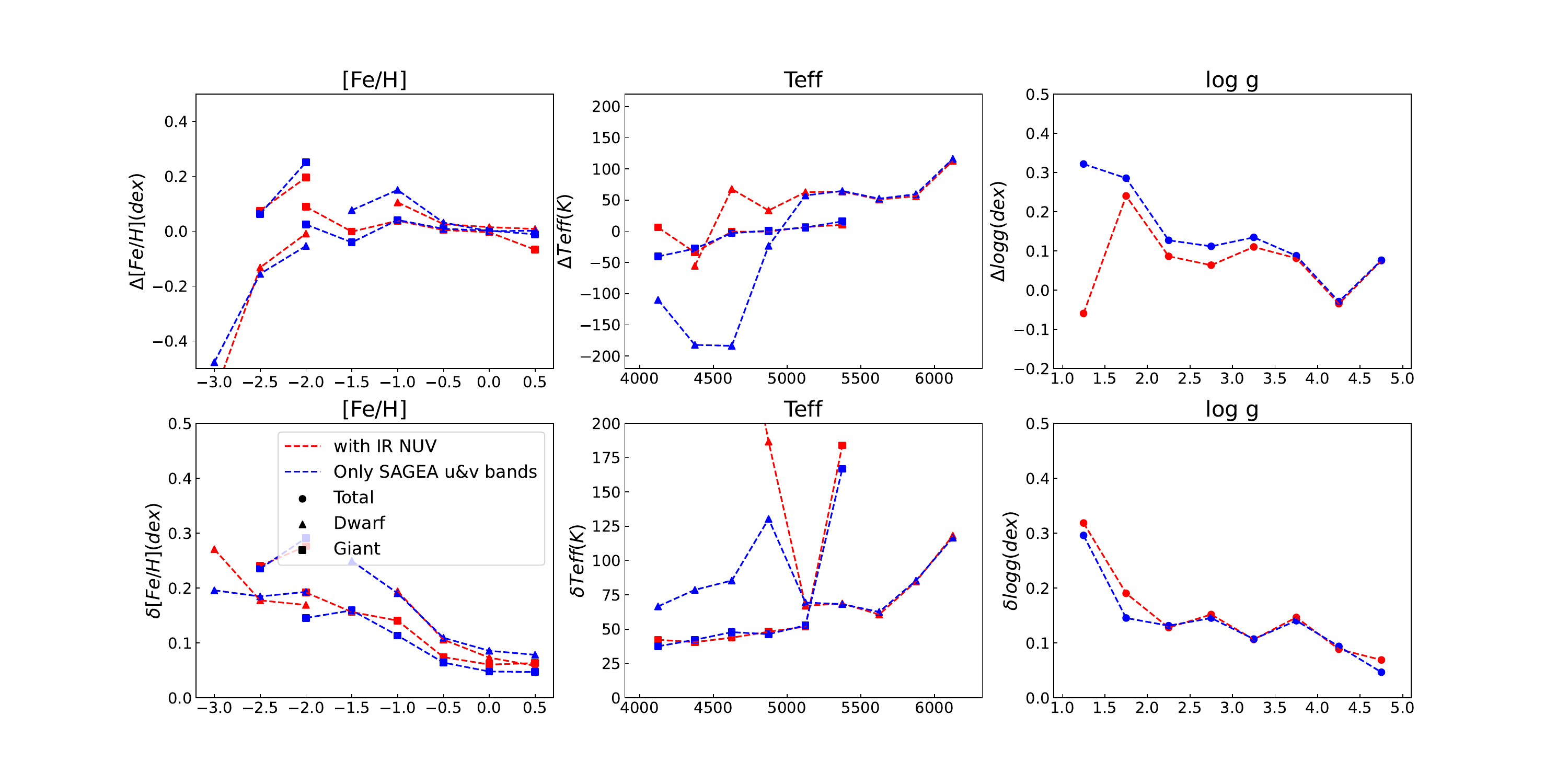}
\caption{Systematic errors and standard deviations for three stellar parameters ([Fe/H], $\it T$$\rm _{eff}$, log $g$), as a function of their values. Different colors represent different test sets; the different shapes of the points represent different types of stars (triangles for dwarfs, squares for giants, circles for giants and dwarfs).}
\end{center}
\end{figure*}

\begin{figure*}
\begin{center}
\includegraphics[scale=0.6,angle=0]{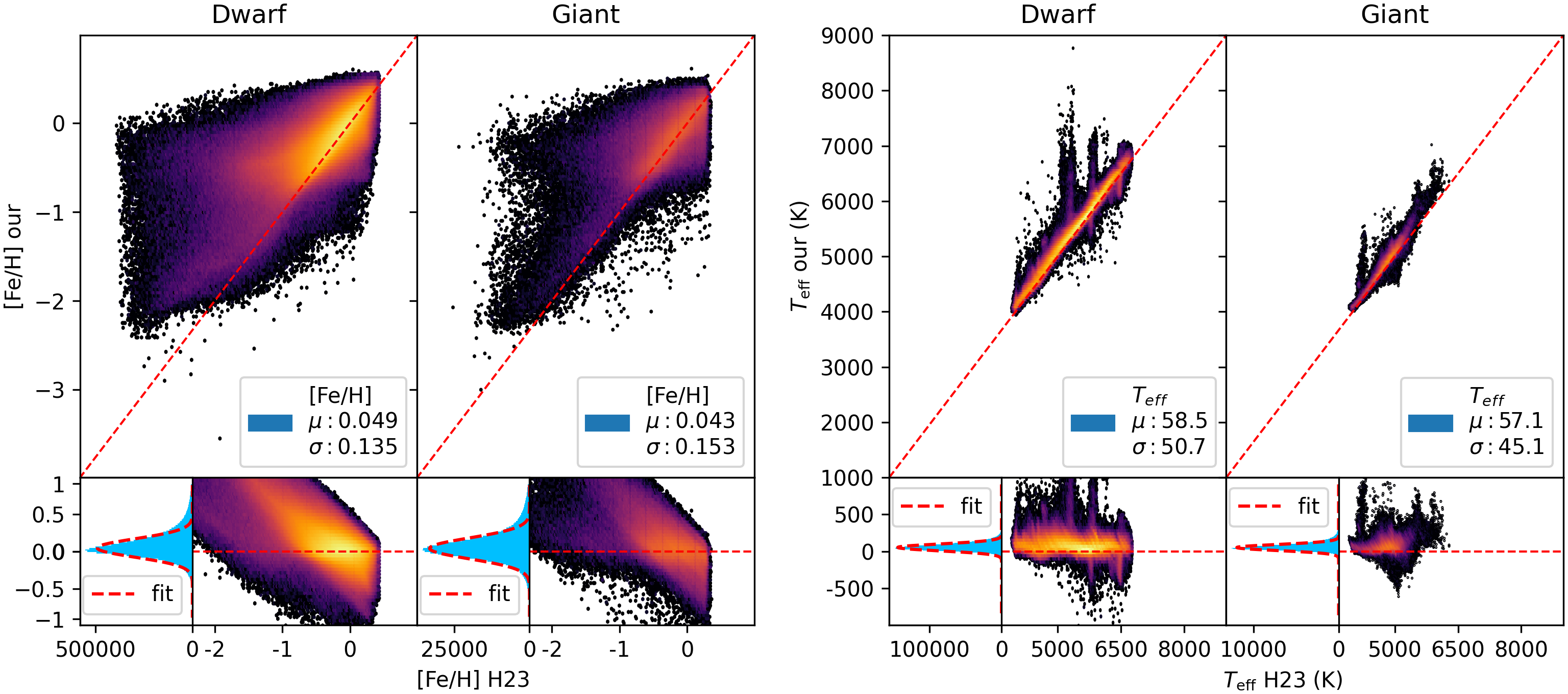}
\caption{A comparison of the performance of the H23 catalog and our catalog for four models with two parameters. The lower half of each panel shows the distribution of the differences between our model and the H23 model, based on the parameters provided by the H23 model, along with a histogram of the distribution of the differences. The mean and standard deviation were obtained through Gaussian fitting of the histogram, and the fitted values are displayed in the legend of each panel. The red-dashed line in the upper panels is the one-to-one line, and is the zero-residual line in the lower panels. }
\end{center}
\end{figure*}

\begin{figure*}
\begin{center}
\includegraphics[scale=0.5,angle=0]{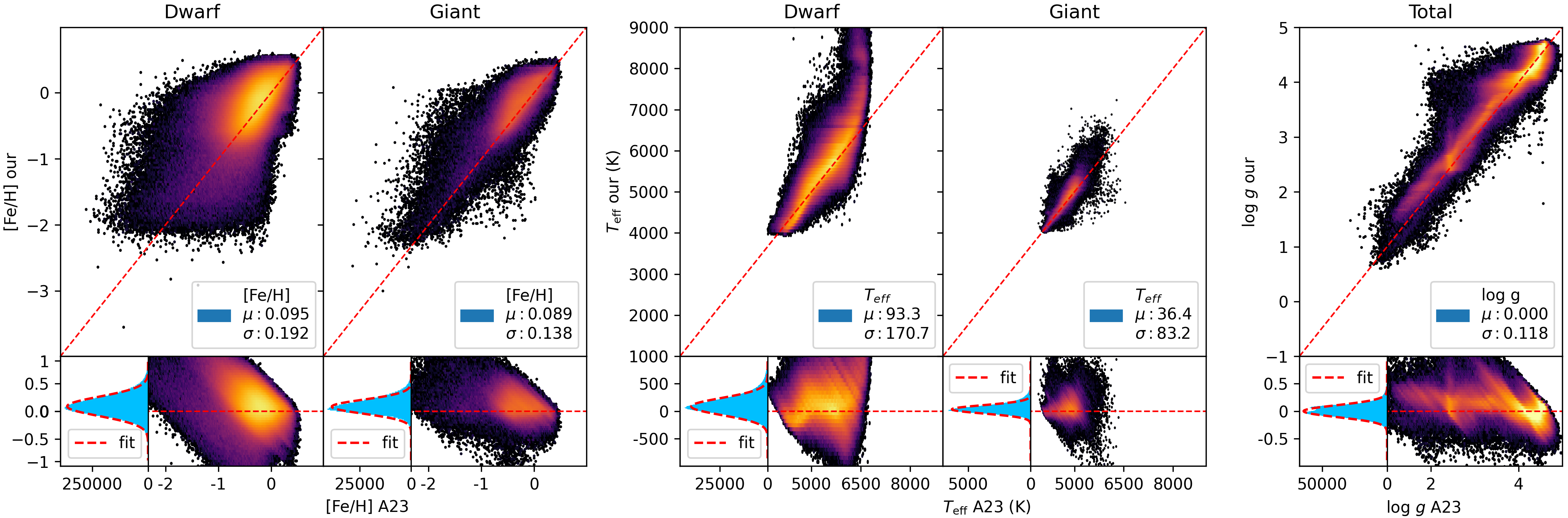}
\caption{Similar to Figure 9, but for a comparison with the A23 catalog.}
\end{center}
\end{figure*}

\vskip 1.5cm
\subsection{Data Set 2: Addition of Two IR Photometric Surveys}

After cross-matching with 2MASS $J$-, $H$-, $K$-bands, WISE $W1$- and $W2$-bands, as well as the GALEX NUV-band, a second set of training sample is assembled containing 178,542 dwarf stars and 11,559 giant stars. 
Figure 7 illustrates the accuracy and precision of the five models on this test set; with each subplot representing the training result of one model. When contrasting Figure 7 with Figure 6, it is clear that the statistical accuracy and precision of the [Fe/H] and log $g$ for the dwarf stars has been improved somewhat after adding the infrared photometric data. However, in subsequent studies, we believe that the higher precision observed here may be an artifact due to differences in sample distribution.

\subsection{Final Accuracy and Precision }

Figure 8 illustrates the performance of the mean and standard deviation for the five models, with the three columns of panels representing the [Fe/H], $\it T$$\rm _{eff}$, and log $g$, respectively. The two rows display the variations of the standard deviation (first row) and mean (second row) with respect to the known stellar parameters. In the figure, different colors represent distinct training sets, while the various shapes represent different types of stars. 

[Fe/H] basically has no systematic error, and the deviation is within 0.1\,dex, but the dispersion is large at the metal-poor end. When [Fe/H] $< -2.0$, it is about 0.2 - 0.3\,dex, and when [Fe/H] $>  -0.5$, it can achieve accuracy better than 0.1\,dex. 
The systematic error in $\it T$$\rm _{eff}$ is approximately 20 K for giant stars and around 50 K for dwarf stars. Outliers are primarily due to the larger intrinsic dispersion caused by small sample sizes and the lower luminosity of cooler stars. For giant stars with temperatures below 5250 K, the temperature dispersion is better than 50 K. For most dwarf stars, the temperature dispersion is around 100 K.
The systematic error of log $g$ is also relatively small as a whole and at about 0.1\,dex. For most stars, the log $g$ accuracy can be better than 0.15\,dex. The dispersion of this parameter becomes larger with decreasing log $g$.

In Figure 8, red-dashed lines represent the systematic error and standard deviation of the corresponding parameters in the test set of Data Set 2. 
When comparing the results of the first dataset (blue-dashed line), it can be seen that the second dataset exhibits similar precision and accuracy across various parameters. However, the second dataset, which includes several different survey data with different limiting magnitudes and a wide range of wavelength, suffers from significant selection effects for both low effective temperature and high effective temperature stars. Additionally, the sample size of the second dataset is only one-tenth that of the first dataset therefore has a smaller coverage range of stellar parameters. In consequence, subsequent tests will utilize only the first dataset.

\subsection{External Tests with Other Catalogs}

In this study, a substantial portion of publicly available spectroscopic survey data was collected for model training and testing. To conduct external tests, we cross-match the stellar parameters obtained from the XP spectra provided by Gaia DR3 (\citealt{2023ApJS..267....8A}, hereafter A23) with those obtained by previous studies utilizing the SAGES survey (\citealt{2023ApJ...957...65H}, hereafter H23).

After cross-matching with the H23 catalog, we identified 13,455,865 sources in common in Data Set 1. The error distribution of the three parameters provided by the five models is depicted in Figure 9. Since the H23 catalog lacks log $g$ information, a direct performance comparison with this model was not possible. The mean and standard deviation of the models were determined by Gaussian fitting of the residual values of stellar parameters measured by the two models (values in our catalog minus values in H23 catalog). The mean offsets for both models in metallicity is close to zero, and the standard deviation is slightly higher than 0.1\,dex, indicating that, in terms of metallicity, the parameters generated by our models are in agreement with the values given in the H23 paper. Assuming equal errors in H23 and our model, the errors for both are approximately 0.1\,dex. 

However, in the effective temperature models, there is a significant systematic difference of about 50 K between the two. This substantial systematic difference may be attributed to the training set. Our training set used the parameters directly provided by LAMOST DR10, while H23 used stellar parameters they calculated themselves from LAMOST spectra. In both the giant and dwarf star temperatures, there are some vertical structures, indicating that a group of stars with similar temperatures in the H23 catalog are considered to have different effective temperature in this work. Given that this structure does not appear in comparisons with other studies, we believe it may be caused by some systematic errors inherent in the SAGES DR1. The polynomial fitting, due to its limited number of parameters, it is unable to correct the complex systematic error in input data therefore cause large systematic error in output data.

Similarly, after cross-matching with the A23 parameter catalog, there are 11,367,335 sources in common. The error distribution of the three parameters provided by the five models is illustrated in Figure 10. The generation method of the figure is consistent with the previous discussion. It can be seen that the overall correlation of the five models is quite good. An unusual structure in the surface gravity and effective temperature models is evident. This may be due to the different training datasets used. In the cross-matched catalog between LAMOST and APOGEE, the log $g$ and $\it T$$\rm _{eff}$ also exhibit similar anomalous structures.
Because the temperature range of the training set used by A23 only extends up to 7000\,K, most stars with temperatures greater than 7000\,K are considered as stars with temperatures of 6500 to 7000\,K. Consequently, a larger standard deviation is exhibited when comparing our model to A23 for dwarf stars.

\subsection{External Tests with Star Clusters}

This section presents the results of the tests using star clusters. We cross-matched our catalog with the proper motion from Gaia DR3, and performed cuts in right ascension and declination near the clusters center, restricting the proper motion errors in both directions to be less than 0.3 mas/yr. We then selected a stringent sample of cluster members by cutting the over-dense regions in the proper motion space. By examining the color-magnitude diagram of this sample, we confirmed that it is virtually free of field star contamination. Using this method for member selection, we identified high-purity member stars for two open clusters, NGC 2420 and NGC 2682, and two globular clusters, NGC 6934 and NGC 7089. After applying a signal-to-noise ratio $> 20$ criterion in the SAGES u- and v- bands, we conducted metallicity precision tests on the selected member stars.

Figure 11 shows the metallicity distributions of the four clusters, fitted with Gaussian functions, and the resulting mean and standard deviation. The standard deviations are consistent with the metallicity dispersions presented in Section 3.3. By comparing the metallicity measurements of the globular and open clusters with those provided by \cite{2010arXiv1012.3224H} (here after H10) and \cite{2018A&A...618A..93C} (here after C18) (indicated by the black dashed lines in Figure 11), it can be seen that the systematic errors in metallicity at the metal-poor end are also consistent with the results from the test set in Section 3.3.

\begin{figure*}
\begin{center}
\includegraphics[scale=0.55,angle=0]{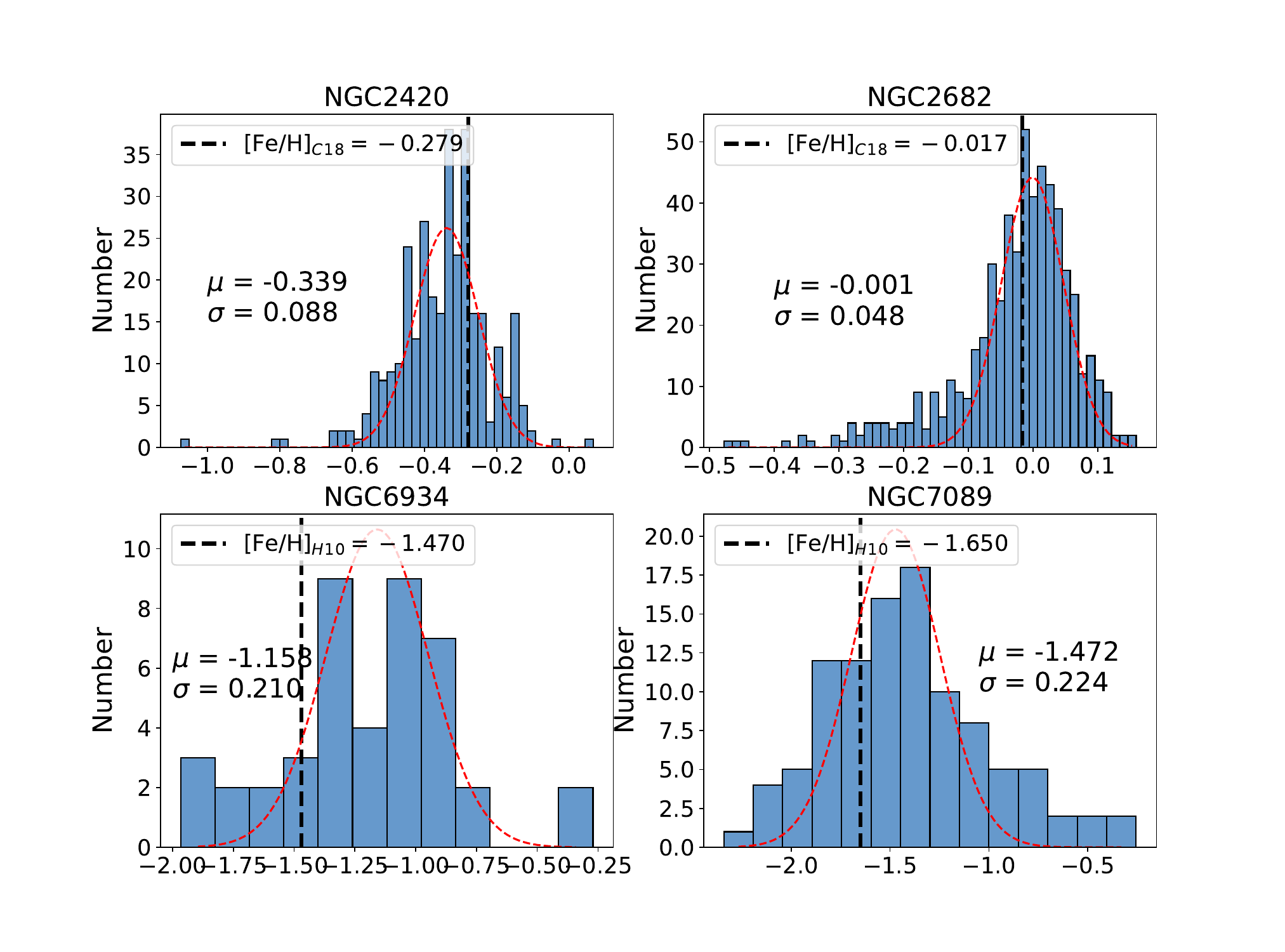}
\caption{This figure presents the metallicity histograms from this work. The top two panels show the open clusters NGC 2420 and NGC 2682, and the bottom two panels display the globular clusters NGC 6934 and NGC 7089. The red dashed lines represent the best-fit Gaussian distributions to the histograms, with the mean and standard deviation displayed in the figures. The black dashed lines indicate the cluster metallicities from literature, with their values shown in legends.
}
\end{center}
\end{figure*}

\begin{figure*}
\begin{center}
\includegraphics[scale=0.55,angle=0]{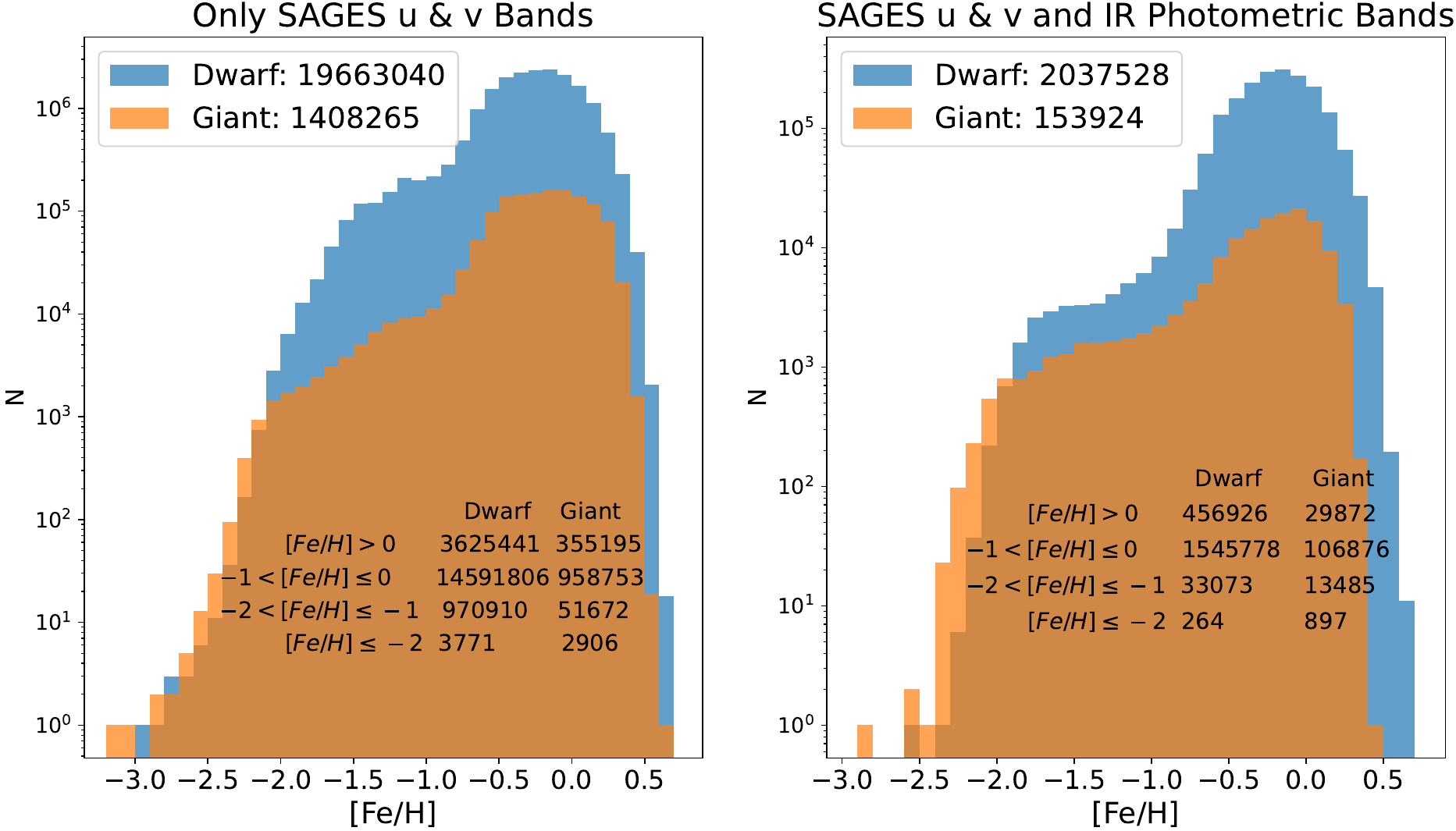}
\caption{{\it \textbf{Left panel:}} The photometric-metallicity distributions of dwarf (blue) and giant (yellow) stars in Data Set 1. The number of dwarf and giant stars in each bin of metallicity are also given. {\it \textbf{Right panel:}} Similar to left panel, but using data from Data Set 2.
}
\end{center}
\end{figure*}

\begin{table*}
\centering
\caption{Description of the Final Sample}
\begin{tabular}{lll}
\hline
\hline
Parameter&Description&Unit\\
\hline
RA&Right Ascension from SAGES DR1 (J2000)&degrees\\
DEC&Declination from SAGES DR1 (J2000)&degrees\\
$\it T$$\rm _{eff}$&Effective temperature (Unreliable data are marked with -9999)&K\\
log $g$&Logarithm surface gravity (Unreliable data are marked with -9999)& cgs\\
$[$Fe/H$]$&Photometric metallicity (Unreliable data are marked with -9999)& \\
ebv\_sfd&Value of E(B-V) from the extinction map of SFD98&--\\
$u/v$&Magnitude for the SAGES $u/v$ bands from DR1&mag\\
$u\_\rm err/v\_\rm err$&Uncentainty of magnitude for the SAGES $u/v$ bands from DR1&mag\\
$G/Bp/Rp$&Magnitudes for three Gaia bands from EDR3&mag\\
$G\_\rm err/Bp\_\rm err/Rp\_\rm err$&Uncentainty of magnitudes for three Gaia bands from EDR3&mag\\
$J/H/K$&Magnitudes for three 2MASS bands&mag\\
$J\_\rm err/H\_\rm err/K\_\rm err$&Uncentainty of magnitudes for three 2MASS bands&mag\\
$W1/W2$&Magnitude for the WISE $W1/W2$ bands&mag\\
$W1\_\rm err/W2\_\rm err$&Uncentainty of magnitude for the WISE $W1/W2$ bands&mag\\
$NUV$&Magnitude for the GALEX $NUV$ bands&mag\\
$NUV\_\rm err$&Uncentainty of magnitude for the GALEX $NUV$ bands&mag\\
parallex&Parallex from Gaia EDR3&mas\\
err\_parallax&Uncertainty of parallax from Gaia EDR3&mas\\
pm/pmra/pmdec&Proper motion in total/Right Ascension/Declination from Gaia EDR3&mas yr$^{-1}$\\
err\_pmra/err\_pmdec&Uncentainty of proper motion in Right Ascension/Declination from Gaia EDR3&mas yr$^{-1}$\\
type&Flag to indicate classifications of stars, 1 means dwarf, 0 means giant&--\\
\hline
\end{tabular}
\end{table*}

\section{The Final Sample}\label{sec4}

Through the application of the trained models to all photometric data, two sets of value added catalogs are obtained. The first data set contains 21,071,305 stars, including 19,663,040 dwarf stars and 1,408,265 giant stars; the second data set has 
2,191,452 stars, including 2,037,528 dwarf stars and 153,924 giant stars.

In view of the poor extrapolation ability of the random forest algorithm, the results obtained from extrapolating Teff were unsatisfactory. To remove this low-temperature star sample, we performed a polynomial fit to the temperature using ($Bp-Rp$)$_0$ from the training data and excluded the three atmospheric parameters of the low-temperature stars with ($Bp-Rp$)$_0 > $ 1.598. Figure 12 shows the distribution of metal abundance predicted by the two models. For the first data set, the fractions of stars with [Fe/H] $\leq 0$, $\leq -1$, and $\leq -2$  are 81.11$\%$, 4.88$\%$, and 0.03$\%$, respectively. For the second data set, the fractions of stars with [Fe/H] $\leq 0$, $\leq -1$, and $\leq -2$  are 77.59$\%$, 2.18$\%$, and 0.05$\%$, respectively. The numbers of stars in each interval of [Fe/H] and stellar types are shown in the histogram. Table 3 lists the parameters provided in the final data product.

\section{Summary}\label{sec5}

In this work, we combine stellar-parameter estimates from spectral surveys and high-resolution spectral data collected from the literature with photometric data from SAGES DR1, Gaia DR3, WISE, 2MASS, and GALEX. Using the random forest algorithm, we derive high-precision stellar parameters for a total of 
21,071,305 stars, including 19,663,040 dwarf stars and 1,408,265 giant stars. The overall precision is about 0.1\,dex in $[$Fe/H$]$ and log $g$, and the $\it T$$\rm _{eff}$ precision is better than 100\,K. 

Currently, the log $g$ of stars primarily relies on the input parallax from Gaia DR3. The ongoing SAGE Survey is observing in DDO51 band, which is sensitive to log g, will provide more accurate log g parameter information once the observations are completed. Subsequently, the large number of stars with available stellar parameters can be used as the basis for a variety of follow-up efforts, such as using metal-abundance information provided by this catalog to identity and study stellar streams with Gaia kinematic information, or to find candidate targets for future spectroscopic observation. At present, two catalogs of this work are available on (will put after accepted) website. 
%This website will be prepared right before accepted

\begin{acknowledgments}
This study is supported by the National Natural Science Foundation of China (NSFC) under grant Nos. 11988101 12261141689, 12090044, 12090040.  T.C.B. acknowledges partial support for this work from grant PHY 14-30152; Physics Frontier Center/JINA Center for the Evolution of the Elements (JINA-CEE), and OISE-1927130: The International Research Network for Nuclear Astrophysics (IReNA), awarded by the US National Science Foundation. We thank the
staff of the University of Arizona and the mountain operation team of the
Steward Observatory, including Bill Wood, Michael Lesser, Ed Olszewski,
Joe Hoscheidt, Gary Rosenbaum, Jeff Rill, and
Richard Green, for assistance with the observations. Sponsored by the Xinjiang Uygur Autonomous Region ‘Tianchi Talent’ Introduction Plan.
%We thank Timothy C. Beers for professional language revision.
\end{acknowledgments}

\vfill\eject

\vfill\eject
\bibliography{paper}{}

\begin{thebibliography}{}
\expandafter\ifx\csname natexlab\endcsname\relax\def\natexlab#1{#1}\fi
\providecommand{\url}[1]{\href{#1}{#1}}
\providecommand{\dodoi}[1]{doi:~\href{http://doi.org/#1}{\nolinkurl{#1}}}
\providecommand{\doeprint}[1]{\href{http://ascl.net/#1}{\nolinkurl{http://ascl.net/#1}}}
\providecommand{\doarXiv}[1]{\href{https://arxiv.org/abs/#1}{\nolinkurl{https://arxiv.org/abs/#1}}}

\bibitem[{{Abdurro'Uf} \& {et al.}(2023)}]{2023yCat.3286....0A}
{Abdurro'Uf}, \& {et al.} 2023, {VizieR Online Data Catalog: APOGEE-2 DR17 final allStar catalog (Abdurro'uf+, 2022)}, VizieR On-line Data Catalog: III/286. Originally published in: 2022ApJS..259...35A

\bibitem[{{An} \& {Beers}(2020)}]{2020ApJ...897...39A}
{An}, D., \& {Beers}, T.~C. 2020, \apj, 897, 39, \dodoi{10.3847/1538-4357/ab8d39}

\bibitem[{{Andrae} {et~al.}(2023){Andrae}, {Rix}, \& {Chandra}}]{2023ApJS..267....8A}
{Andrae}, R., {Rix}, H.-W., \& {Chandra}, V. 2023, \apjs, 267, 8, \dodoi{10.3847/1538-4365/acd53e}

\bibitem[{{Aoki} {et~al.}(2013){Aoki}, {Beers}, {Lee}, {Honda}, {Ito}, {Takada-Hidai}, {Frebel}, {Suda}, {Fujimoto}, {Carollo}, \& {Sivarani}}]{2013AJ....145...13A}
{Aoki}, W., {Beers}, T.~C., {Lee}, Y.~S., {et~al.} 2013, \aj, 145, 13, \dodoi{10.1088/0004-6256/145/1/13}

\bibitem[{{Bai} {et~al.}(2019){Bai}, {Liu}, {Bai}, {Wang}, \& {Fan}}]{2019AJ....158...93B}
{Bai}, Y., {Liu}, J., {Bai}, Z., {Wang}, S., \& {Fan}, D. 2019, \aj, 158, 93, \dodoi{10.3847/1538-3881/ab3048}

\bibitem[{{Bianchi} {et~al.}(2014){Bianchi}, {Conti}, \& {Shiao}}]{2014yCat.2335....0B}
{Bianchi}, L., {Conti}, A., \& {Shiao}, B. 2014, {VizieR Online Data Catalog: GALEX-GR6/7 data release (Bianchi+ 2014)}, VizieR On-line Data Catalog: II/335. Originally published in: 2014AdSpR..53..900B

\bibitem[{{Bonifacio} {et~al.}(2009){Bonifacio}, {Spite}, {Cayrel}, {Hill}, {Spite}, {Fran{\c{c}}ois}, {Plez}, {Ludwig}, {Caffau}, {Molaro}, {Depagne}, {Andersen}, {Barbuy}, {Beers}, {Nordstr{\"o}m}, \& {Primas}}]{2009A&A...501..519B}
{Bonifacio}, P., {Spite}, M., {Cayrel}, R., {et~al.} 2009, \aap, 501, 519, \dodoi{10.1051/0004-6361/200810610}

\bibitem[{{Bonoli} {et~al.}(2021){Bonoli}, {Mar{\'\i}n-Franch}, {Varela}, {V{\'a}zquez Rami{\'o}}, {Abramo}, {Cenarro}, {Dupke}, {V{\'\i}lchez}, {Crist{\'o}bal-Hornillos}, {Gonz{\'a}lez Delgado}, {Hern{\'a}ndez-Monteagudo}, {L{\'o}pez-Sanjuan}, {Muniesa}, {Civera}, {Ederoclite}, {Hern{\'a}n-Caballero}, {Marra}, {Baqui}, {Cortesi}, {Cypriano}, {Daflon}, {de Amorim}, {D{\'\i}az-Garc{\'\i}a}, {Diego}, {Mart{\'\i}nez-Solaeche}, {P{\'e}rez}, {Placco}, {Prada}, {Queiroz}, {Alcaniz}, {Alvarez-Candal}, {Cepa}, {Maroto}, {Roig}, {Siffert}, {Taylor}, {Benitez}, {Moles}, {Sodr{\'e}}, {Carneiro}, {Mendes de Oliveira}, {Abdalla}, {Angulo}, {Aparicio Resco}, {Balaguera-Antol{\'\i}nez}, {Ballesteros}, {Brito-Silva}, {Broadhurst}, {Carrasco}, {Castro}, {Cid Fernandes}, {Coelho}, {de Melo}, {Doubrawa}, {Fernandez-Soto}, {Ferrari}, {Finoguenov}, {Garc{\'\i}a-Benito}, {Iglesias-P{\'a}ramo}, {Jim{\'e}nez-Teja}, {Kitaura}, {Laur}, {Lopes}, {Lucatelli}, {Mart{\'\i}nez}, {Maturi}, {Overzier}, {Pigozzo}, {Quartin}, {Rodr{\'\i}guez-Mart{\'\i}n}, {Salzano}, {Tamm}, {Tempel}, {Umetsu}, {Valdivielso}, {von Marttens}, {Zitrin}, {D{\'\i}az-Mart{\'\i}n}, {L{\'o}pez-Alegre}, {L{\'o}pez-Sainz}, {Yanes-D{\'\i}az}, {Rueda-Teruel}, {Rueda-Teruel}, {Abril Iba{\~n}ez}, {L Ant{\'o}n Bravo}, {Bello Ferrer}, {Bielsa}, {Casino}, {Castillo}, {Chueca}, {Cuesta}, {Garzar{\'a}n Calderaro}, {Iglesias-Marzoa}, {{\'I}niguez}, {Lamadrid Gutierrez}, {Lopez-Martinez}, {Lozano-P{\'e}rez}, {Ma{\'\i}cas Sacrist{\'a}n}, {Molina-Ib{\'a}{\~n}ez}, {Moreno-Signes}, {Rodr{\'\i}guez Llano}, {Royo Navarro}, {Tilve Rua}, {Andrade}, {Alfaro}, {Akras}, {Arnalte-Mur}, {Ascaso}, {Barbosa}, {Beltr{\'a}n Jim{\'e}nez}, {Benetti}, {Bengaly}, {Bernui}, {Blanco-Pillado}, {Borges Fernandes}, {Bregman}, {Bruzual}, {Calderone}, {Carvano}, {Casarini}, {Chaves-Montero}, {Chies-Santos}, {Coutinho de Carvalho}, {Dimauro}, {Duarte Puertas}, {Figueruelo}, {Gonz{\'a}lez-Serrano}, {Guerrero}, {Gurung-L{\'o}pez}, {Herranz}, {Huertas-Company}, {Irwin}, {Izquierdo-Villalba}, {Kanaan}, {Kehrig}, {Kirkpatrick}, {Lim}, {Lopes}, {Lopes de Oliveira}, {Marcos-Caballero}, {Mart{\'\i}nez-Delgado}, {Mart{\'\i}nez-Gonz{\'a}lez}, {Mart{\'\i}nez-Somonte}, {Oliveira}, {Orsi}, {Penna-Lima}, {Reis}, {Spinoso}, {Tsujikawa}, {Vielva}, {Vitorelli}, {Xia}, {Yuan}, {Arroyo-Polonio}, {Dantas}, {Galarza}, {Gon{\c{c}}alves}, {Gon{\c{c}}alves}, {Gonzalez}, {Gonzalez}, {Greisel}, {Jim{\'e}nez-Esteban}, {Landim}, {Lazzaro}, {Magris}, {Monteiro-Oliveira}, {Pereira}, {Rebou{\c{c}}as}, {Rodriguez-Espinosa}, {Santos da Costa}, \& {Telles}}]{2021A&A...653A..31B}
{Bonoli}, S., {Mar{\'\i}n-Franch}, A., {Varela}, J., {et~al.} 2021, \aap, 653, A31, \dodoi{10.1051/0004-6361/202038841}

\bibitem[{{Breiman}(2001)}]{2001MachL..45....5B}
{Breiman}, L. 2001, Machine Learning, 45, 5, \dodoi{10.1023/A:1010933404324}

\bibitem[{{Cantat-Gaudin} {et~al.}(2018){Cantat-Gaudin}, {Jordi}, {Vallenari}, {Bragaglia}, {Balaguer-N{\'u}{\~n}ez}, {Soubiran}, {Bossini}, {Moitinho}, {Castro-Ginard}, {Krone-Martins}, {Casamiquela}, {Sordo}, \& {Carrera}}]{2018A&A...618A..93C}
{Cantat-Gaudin}, T., {Jordi}, C., {Vallenari}, A., {et~al.} 2018, \aap, 618, A93, \dodoi{10.1051/0004-6361/201833476}

\bibitem[{{Cenarro} {et~al.}(2019){Cenarro}, {Moles}, {Crist{\'o}bal-Hornillos}, {Mar{\'\i}n-Franch}, {Ederoclite}, {Varela}, {L{\'o}pez-Sanjuan}, {Hern{\'a}ndez-Monteagudo}, {Angulo}, {V{\'a}zquez Rami{\'o}}, {Viironen}, {Bonoli}, {Orsi}, {Hurier}, {San Roman}, {Greisel}, {Vilella-Rojo}, {D{\'\i}az-Garc{\'\i}a}, {Logro{\~n}o-Garc{\'\i}a}, {Gurung-L{\'o}pez}, {Spinoso}, {Izquierdo-Villalba}, {Aguerri}, {Allende Prieto}, {Bonatto}, {Carvano}, {Chies-Santos}, {Daflon}, {Dupke}, {Falc{\'o}n-Barroso}, {Gon{\c{c}}alves}, {Jim{\'e}nez-Teja}, {Molino}, {Placco}, {Solano}, {Whitten}, {Abril}, {Ant{\'o}n}, {Bello}, {Bielsa de Toledo}, {Castillo-Ram{\'\i}rez}, {Chueca}, {Civera}, {D{\'\i}az-Mart{\'\i}n}, {Dom{\'\i}nguez-Mart{\'\i}nez}, {Garzar{\'a}n-Calderaro}, {Hern{\'a}ndez-Fuertes}, {Iglesias-Marzoa}, {I{\~n}iguez}, {Jim{\'e}nez Ruiz}, {Kruuse}, {Lamadrid}, {Lasso-Cabrera}, {L{\'o}pez-Alegre}, {L{\'o}pez-Sainz}, {Ma{\'\i}cas}, {Moreno-Signes}, {Muniesa}, {Rodr{\'\i}guez-Llano}, {Rueda-Teruel}, {Rueda-Teruel}, {Soriano-Lagu{\'\i}a}, {Tilve}, {Valdivielso}, {Yanes-D{\'\i}az}, {Alcaniz}, {Mendes de Oliveira}, {Sodr{\'e}}, {Coelho}, {Lopes de Oliveira}, {Tamm}, {Xavier}, {Abramo}, {Akras}, {Alfaro}, {Alvarez-Candal}, {Ascaso}, {Beasley}, {Beers}, {Borges Fernandes}, {Bruzual}, {Buzzo}, {Carrasco}, {Cepa}, {Cortesi}, {Costa-Duarte}, {De Pr{\'a}}, {Favole}, {Galarza}, {Galbany}, {Garcia}, {Gonz{\'a}lez Delgado}, {Gonz{\'a}lez-Serrano}, {Guti{\'e}rrez-Soto}, {Hernandez-Jimenez}, {Kanaan}, {Kuncarayakti}, {Landim}, {Laur}, {Licandro}, {Lima Neto}, {Lyman}, {Ma{\'\i}z Apell{\'a}niz}, {Miralda-Escud{\'e}}, {Morate}, {Nogueira-Cavalcante}, {Novais}, {Oncins}, {Oteo}, {Overzier}, {Pereira}, {Rebassa-Mansergas}, {Reis}, {Roig}, {Sako}, {Salvador-Rusi{\~n}ol}, {Sampedro}, {S{\'a}nchez-Bl{\'a}zquez}, {Santos}, {Schmidtobreick}, {Siffert}, {Telles}, \& {Vilchez}}]{2019A&A...622A.176C}
{Cenarro}, A.~J., {Moles}, M., {Crist{\'o}bal-Hornillos}, D., {et~al.} 2019, \aap, 622, A176, \dodoi{10.1051/0004-6361/201833036}

\bibitem[{{Chiti} {et~al.}(2021){Chiti}, {Frebel}, {Mardini}, {Daniel}, {Ou}, \& {Uvarova}}]{2021ApJS..254...31C}
{Chiti}, A., {Frebel}, A., {Mardini}, M.~K., {et~al.} 2021, \apjs, 254, 31, \dodoi{10.3847/1538-4365/abf73d}

\bibitem[{{Crawford} {et~al.}(1970){Crawford}, {Barnes}, \& {Golson}}]{1970AJ.....75..624C}
{Crawford}, D.~L., {Barnes}, J.~V., \& {Golson}, J.~C. 1970, \aj, 75, 624, \dodoi{10.1086/110996}

\bibitem[{{Cui} {et~al.}(2012){Cui}, {Zhao}, {Chu}, {Li}, {Li}, {Zhang}, {Su}, {Yao}, {Wang}, {Xing}, {Li}, {Zhu}, {Wang}, {Gu}, {Luo}, {Xu}, {Zhang}, {Liu}, {Zhang}, {Yang}, {Cao}, {Chen}, {Chen}, {Chen}, {Chen}, {Chu}, {Feng}, {Gong}, {Hou}, {Hu}, {Hu}, {Hu}, {Jia}, {Jiang}, {Jiang}, {Jiang}, {Jin}, {Li}, {Li}, {Li}, {Liu}, {Liu}, {Lu}, {Mao}, {Men}, {Qi}, {Qi}, {Shi}, {Tang}, {Tao}, {Wang}, {Wang}, {Wang}, {Wang}, {Wang}, {Wang}, {Wang}, {Wang}, {Wang}, {Wang}, {Wang}, {Wang}, {Xu}, {Xu}, {Yang}, {Yu}, {Yuan}, {Yuan}, {Zhai}, {Zhang}, {Zhang}, {Zhang}, {Zhao}, {Zhou}, {Zhou}, {Zhu}, \& {Zou}}]{2012RAA....12.1197C}
{Cui}, X.-Q., {Zhao}, Y.-H., {Chu}, Y.-Q., {et~al.} 2012, Research in Astronomy and Astrophysics, 12, 1197, \dodoi{10.1088/1674-4527/12/9/003}

\bibitem[{{Cutri} {et~al.}(2003){Cutri}, {Skrutskie}, {van Dyk}, {Beichman}, {Carpenter}, {Chester}, {Cambresy}, {Evans}, {Fowler}, {Gizis}, {Howard}, {Huchra}, {Jarrett}, {Kopan}, {Kirkpatrick}, {Light}, {Marsh}, {McCallon}, {Schneider}, {Stiening}, {Sykes}, {Weinberg}, {Wheaton}, {Wheelock}, \& {Zacarias}}]{2003tmc..book.....C}
{Cutri}, R.~M., {Skrutskie}, M.~F., {van Dyk}, S., {et~al.} 2003, {2MASS All Sky Catalog of point sources.}

\bibitem[{{Cutri} {et~al.}(2021){Cutri}, {Wright}, {Conrow}, {Fowler}, {Eisenhardt}, {Grillmair}, {Kirkpatrick}, {Masci}, {McCallon}, {Wheelock}, {Fajardo-Acosta}, {Yan}, {Benford}, {Harbut}, {Jarrett}, {Lake}, {Leisawitz}, {Ressler}, {Stanford}, {Tsai}, {Liu}, {Helou}, {Mainzer}, {Gettngs}, {Gonzalez}, {Hoffman}, {Marsh}, {Padgett}, {Skrutskie}, {Beck}, {Papin}, \& {Wittman}}]{2014yCat.2328....0C}
{Cutri}, R.~M., {Wright}, E.~L., {Conrow}, T., {et~al.} 2021, VizieR Online Data Catalog, II/328

\bibitem[{{Ezzeddine} {et~al.}(2020){Ezzeddine}, {Rasmussen}, {Frebel}, {Chiti}, {Hinojisa}, {Placco}, {Ji}, {Beers}, {Hansen}, {Roederer}, {Sakari}, \& {Melendez}}]{2020ApJ...898..150E}
{Ezzeddine}, R., {Rasmussen}, K., {Frebel}, A., {et~al.} 2020, \apj, 898, 150, \dodoi{10.3847/1538-4357/ab9d1a}

\bibitem[{{Fan} {et~al.}(2023){Fan}, {Zhao}, {Wang}, {Zheng}, {Zhao}, {Li}, {Chen}, {Yuan}, {Li}, {Tan}, {Song}, {Zuo}, {Huang}, {Luo}, {Esamdin}, {Ma}, {Li}, {Song}, {Grupp}, {Zhao}, {Ehgamberdiev}, {Burkhonov}, {Feng}, {Bai}, {Zhang}, {Niu}, {Khodjaev}, {Khafizov}, {Asfandiyarov}, {Shaymanov}, {Karimov}, {Yuldashev}, {Lu}, {Zhaori}, {Hong}, {Hu}, {Liu}, \& {Xu}}]{2023ApJS..268....9F}
{Fan}, Z., {Zhao}, G., {Wang}, W., {et~al.} 2023, \apjs, 268, 9, \dodoi{10.3847/1538-4365/ace04a}

\bibitem[{{Fran{\c{c}}ois} {et~al.}(2018){Fran{\c{c}}ois}, {Caffau}, {Bonifacio}, {Spite}, {Spite}, {Cayrel}, {Christlieb}, {Gallagher}, {Klessen}, {Koch}, {Ludwig}, {Monaco}, {Plez}, {Steffen}, \& {Zaggia}}]{2018A&A...620A.187F}
{Fran{\c{c}}ois}, P., {Caffau}, E., {Bonifacio}, P., {et~al.} 2018, \aap, 620, A187, \dodoi{10.1051/0004-6361/201834375}

\bibitem[{{Gaia Collaboration} {et~al.}(2021){Gaia Collaboration}, {Brown}, {Vallenari}, {Prusti}, {de Bruijne}, {Babusiaux}, {Biermann}, {Creevey}, {Evans}, {Eyer}, {Hutton}, {Jansen}, {Jordi}, {Klioner}, {Lammers}, {Lindegren}, {Luri}, {Mignard}, {Panem}, {Pourbaix}, {Randich}, {Sartoretti}, {Soubiran}, {Walton}, {Arenou}, {Bailer-Jones}, {Bastian}, {Cropper}, {Drimmel}, {Katz}, {Lattanzi}, {van Leeuwen}, {Bakker}, {Cacciari}, {Casta{\~n}eda}, {De Angeli}, {Ducourant}, {Fabricius}, {Fouesneau}, {Fr{\'e}mat}, {Guerra}, {Guerrier}, {Guiraud}, {Jean-Antoine Piccolo}, {Masana}, {Messineo}, {Mowlavi}, {Nicolas}, {Nienartowicz}, {Pailler}, {Panuzzo}, {Riclet}, {Roux}, {Seabroke}, {Sordo}, {Tanga}, {Th{\'e}venin}, {Gracia-Abril}, {Portell}, {Teyssier}, {Altmann}, {Andrae}, {Bellas-Velidis}, {Benson}, {Berthier}, {Blomme}, {Brugaletta}, {Burgess}, {Busso}, {Carry}, {Cellino}, {Cheek}, {Clementini}, {Damerdji}, {Davidson}, {Delchambre}, {Dell'Oro}, {Fern{\'a}ndez-Hern{\'a}ndez}, {Galluccio}, {Garc{\'\i}a-Lario}, {Garcia-Reinaldos}, {Gonz{\'a}lez-N{\'u}{\~n}ez}, {Gosset}, {Haigron}, {Halbwachs}, {Hambly}, {Harrison}, {Hatzidimitriou}, {Heiter}, {Hern{\'a}ndez}, {Hestroffer}, {Hodgkin}, {Holl}, {Jan{\ss}en}, {Jevardat de Fombelle}, {Jordan}, {Krone-Martins}, {Lanzafame}, {L{\"o}ffler}, {Lorca}, {Manteiga}, {Marchal}, {Marrese}, {Moitinho}, {Mora}, {Muinonen}, {Osborne}, {Pancino}, {Pauwels}, {Petit}, {Recio-Blanco}, {Richards}, {Riello}, {Rimoldini}, {Robin}, {Roegiers}, {Rybizki}, {Sarro}, {Siopis}, {Smith}, {Sozzetti}, {Ulla}, {Utrilla}, {van Leeuwen}, {van Reeven}, {Abbas}, {Abreu Aramburu}, {Accart}, {Aerts}, {Aguado}, {Ajaj}, {Altavilla}, {{\'A}lvarez}, {{\'A}lvarez Cid-Fuentes}, {Alves}, {Anderson}, {Anglada Varela}, {Antoja}, {Audard}, {Baines}, {Baker}, {Balaguer-N{\'u}{\~n}ez}, {Balbinot}, {Balog}, {Barache}, {Barbato}, {Barros}, {Barstow}, {Bartolom{\'e}}, {Bassilana}, {Bauchet}, {Baudesson-Stella}, {Becciani}, {Bellazzini}, {Bernet}, {Bertone}, {Bianchi}, {Blanco-Cuaresma}, {Boch}, {Bombrun}, {Bossini}, {Bouquillon}, {Bragaglia}, {Bramante}, {Breedt}, {Bressan}, {Brouillet}, {Bucciarelli}, {Burlacu}, {Busonero}, {Butkevich}, {Buzzi}, {Caffau}, {Cancelliere}, {C{\'a}novas}, {Cantat-Gaudin}, {Carballo}, {Carlucci}, {Carnerero}, {Carrasco}, {Casamiquela}, {Castellani}, {Castro-Ginard}, {Castro Sampol}, {Chaoul}, {Charlot}, {Chemin}, {Chiavassa}, {Cioni}, {Comoretto}, {Cooper}, {Cornez}, {Cowell}, {Crifo}, {Crosta}, {Crowley}, {Dafonte}, {Dapergolas}, {David}, {David}, {de Laverny}, {De Luise}, {De March}, {De Ridder}, {de Souza}, {de Teodoro}, {de Torres}, {del Peloso}, {del Pozo}, {Delbo}, {Delgado}, {Delgado}, {Delisle}, {Di Matteo}, {Diakite}, {Diener}, {Distefano}, {Dolding}, {Eappachen}, {Edvardsson}, {Enke}, {Esquej}, {Fabre}, {Fabrizio}, {Faigler}, {Fedorets}, {Fernique}, {Fienga}, {Figueras}, {Fouron}, {Fragkoudi}, {Fraile}, {Franke}, {Gai}, {Garabato}, {Garcia-Gutierrez}, {Garc{\'\i}a-Torres}, {Garofalo}, {Gavras}, {Gerlach}, {Geyer}, {Giacobbe}, {Gilmore}, {Girona}, {Giuffrida}, {Gomel}, {Gomez}, {Gonzalez-Santamaria}, {Gonz{\'a}lez-Vidal}, {Granvik}, {Guti{\'e}rrez-S{\'a}nchez}, {Guy}, {Hauser}, {Haywood}, {Helmi}, {Hidalgo}, {Hilger}, {H{\l}adczuk}, {Hobbs}, {Holland}, {Huckle}, {Jasniewicz}, {Jonker}, {Juaristi Campillo}, {Julbe}, {Karbevska}, {Kervella}, {Khanna}, {Kochoska}, {Kontizas}, {Kordopatis}, {Korn}, {Kostrzewa-Rutkowska}, {Kruszy{\'n}ska}, {Lambert}, {Lanza}, {Lasne}, {Le Campion}, {Le Fustec}, {Lebreton}, {Lebzelter}, {Leccia}, {Leclerc}, {Lecoeur-Taibi}, {Liao}, {Licata}, {Lindstr{\o}m}, {Lister}, {Livanou}, {Lobel}, {Madrero Pardo}, {Managau}, {Mann}, {Marchant}, {Marconi}, {Marcos Santos}, {Marinoni}, {Marocco}, {Marshall}, {Martin Polo}, {Mart{\'\i}n-Fleitas}, {Masip}, {Massari}, {Mastrobuono-Battisti}, {Mazeh}, {McMillan}, {Messina}, {Michalik}, {Millar}, {Mints}, {Molina}, {Molinaro}, {Moln{\'a}r}, {Montegriffo}, {Mor}, {Morbidelli}, {Morel}, {Morris}, {Mulone}, {Munoz}, {Muraveva}, {Murphy}, {Musella}, {Noval}, {Ord{\'e}novic}, {Orr{\`u}}, {Osinde}, {Pagani}, {Pagano}, {Palaversa}, {Palicio}, {Panahi}, {Pawlak}, {Pe{\~n}alosa Esteller}, {Penttil{\"a}}, {Piersimoni}, {Pineau}, {Plachy}, {Plum}, {Poggio}, {Poretti}, {Poujoulet}, {Pr{\v{s}}a}, {Pulone}, {Racero}, {Ragaini}, {Rainer}, {Raiteri}, {Rambaux}, {Ramos}, {Ramos-Lerate}, {Re Fiorentin}, {Regibo}, {Reyl{\'e}}, {Ripepi}, {Riva}, {Rixon}, {Robichon}, {Robin}, {Roelens}, {Rohrbasser}, {Romero-G{\'o}mez}, {Rowell}, {Royer}, {Rybicki}, {Sadowski}, {Sagrist{\`a} Sell{\'e}s}, {Sahlmann}, {Salgado}, {Salguero}, {Samaras}, {Sanchez Gimenez}, {Sanna}, {Santove{\~n}a}, {Sarasso}, {Schultheis}, {Sciacca}, {Segol}, {Segovia}, {S{\'e}gransan}, {Semeux}, {Shahaf}, {Siddiqui}, {Siebert}, {Siltala}, {Slezak}, {Smart}, {Solano}, {Solitro}, {Souami}, {Souchay}, {Spagna}, {Spoto}, {Steele}, {Steidelm{\"u}ller}, {Stephenson}, {S{\"u}veges}, {Szabados}, {Szegedi-Elek}, {Taris}, {Tauran}, {Taylor}, {Teixeira}, {Thuillot}, {Tonello}, {Torra}, {Torra}, {Turon}, {Unger}, {Vaillant}, {van Dillen}, {Vanel}, {Vecchiato}, {Viala}, {Vicente}, {Voutsinas}, {Weiler}, {Wevers}, {Wyrzykowski}, {Yoldas}, {Yvard}, {Zhao}, {Zorec}, {Zucker}, {Zurbach}, \& {Zwitter}}]{2021A&A...649A...1G}
{Gaia Collaboration}, {Brown}, A.~G.~A., {Vallenari}, A., {et~al.} 2021, \aap, 649, A1, \dodoi{10.1051/0004-6361/202039657}

\bibitem[{{Galarza} {et~al.}(2022){Galarza}, {Daflon}, {Placco}, {Allende Prieto}, {Borges Fernandes}, {Yuan}, {L{\'o}pez-Sanjuan}, {Lee}, {Solano}, {Jim{\'e}nez-Esteban}, {Sobral}, {Alvarez Candal}, {Pereira}, {Akras}, {Mart{\'\i}n}, {Jim{\'e}nez Teja}, {Cenarro}, {Crist{\'o}bal-Hornillos}, {Hern{\'a}ndez-Monteagudo}, {Mar{\'\i}n-Franch}, {Moles}, {Varela}, {V{\'a}zquez Rami{\'o}}, {Alcaniz}, {Dupke}, {Ederoclite}, {Sodr{\'e}}, \& {Angulo}}]{2022A&A...657A..35G}
{Galarza}, C.~A., {Daflon}, S., {Placco}, V.~M., {et~al.} 2022, \aap, 657, A35, \dodoi{10.1051/0004-6361/202141717}

\bibitem[{{Harris}(2010)}]{2010arXiv1012.3224H}
{Harris}, W.~E. 2010, arXiv e-prints, arXiv:1012.3224, \dodoi{10.48550/arXiv.1012.3224}

\bibitem[{{Hauck} \& {Mermilliod}(1998)}]{1998A&AS..129..431H}
{Hauck}, B., \& {Mermilliod}, M. 1998, \aaps, 129, 431, \dodoi{10.1051/aas:1998195}

\bibitem[{{Huang} {et~al.}(2019){Huang}, {Chen}, {Yuan}, {Zhang}, {Xiang}, {Wang}, {Wang}, {Wolf}, {Liu}, \& {Liu}}]{2019ApJS..243....7H}
{Huang}, Y., {Chen}, B.~Q., {Yuan}, H.~B., {et~al.} 2019, \apjs, 243, 7, \dodoi{10.3847/1538-4365/ab1f72}

\bibitem[{{Huang} {et~al.}(2022){Huang}, {Beers}, {Wolf}, {Lee}, {Onken}, {Yuan}, {Shank}, {Zhang}, {Wang}, {Shi}, \& {Fan}}]{2022ApJ...925..164H}
{Huang}, Y., {Beers}, T.~C., {Wolf}, C., {et~al.} 2022, \apj, 925, 164, \dodoi{10.3847/1538-4357/ac21cb}

\bibitem[{{Huang} {et~al.}(2023){Huang}, {Beers}, {Yuan}, {Tan}, {Wang}, {Zheng}, {Li}, {Lee}, {Li}, {Zhao}, {Xue}, {Liu}, {Zhang}, {Sun}, {Li}, {Gu}, {Wolf}, {Onken}, {Liu}, {Fan}, \& {Zhao}}]{2023ApJ...957...65H}
{Huang}, Y., {Beers}, T.~C., {Yuan}, H., {et~al.} 2023, \apj, 957, 65, \dodoi{10.3847/1538-4357/ace628}

\bibitem[{{Ivezi{\'c}} {et~al.}(2008){Ivezi{\'c}}, {Sesar}, {Juri{\'c}}, {Bond}, {Dalcanton}, {Rockosi}, {Yanny}, {Newberg}, {Beers}, {Allende Prieto}, {Wilhelm}, {Lee}, {Sivarani}, {Norris}, {Bailer-Jones}, {Re Fiorentin}, {Schlegel}, {Uomoto}, {Lupton}, {Knapp}, {Gunn}, {Covey}, {Allyn Smith}, {Miknaitis}, {Doi}, {Tanaka}, {Fukugita}, {Kent}, {Finkbeiner}, {Munn}, {Pier}, {Quinn}, {Hawley}, {Anderson}, {Kiuchi}, {Chen}, {Bushong}, {Sohi}, {Haggard}, {Kimball}, {Barentine}, {Brewington}, {Harvanek}, {Kleinman}, {Krzesinski}, {Long}, {Nitta}, {Snedden}, {Lee}, {Harris}, {Brinkmann}, {Schneider}, \& {York}}]{2008ApJ...684..287I}
{Ivezi{\'c}}, {\v{Z}}., {Sesar}, B., {Juri{\'c}}, M., {et~al.} 2008, \apj, 684, 287, \dodoi{10.1086/589678}

\bibitem[{{Keller} {et~al.}(2007){Keller}, {Bessell}, {Schmidt}, \& {Francis}}]{2007ASPC..364..177K}
{Keller}, S., {Bessell}, M., {Schmidt}, B., \& {Francis}, P. 2007, in Astronomical Society of the Pacific Conference Series, Vol. 364, The Future of Photometric, Spectrophotometric and Polarimetric Standardization, ed. C.~{Sterken}, 177

\bibitem[{{Ksoll} {et~al.}(2020){Ksoll}, {Ardizzone}, {Klessen}, {Koethe}, {Sabbi}, {Robberto}, {Gouliermis}, {Rother}, {Zeidler}, \& {Gennaro}}]{2020MNRAS.499.5447K}
{Ksoll}, V.~F., {Ardizzone}, L., {Klessen}, R., {et~al.} 2020, \mnras, 499, 5447, \dodoi{10.1093/mnras/staa2931}

\bibitem[{{Kunder} {et~al.}(2017){Kunder}, {Kordopatis}, {Steinmetz}, {Zwitter}, {McMillan}, {Casagrande}, {Enke}, {Wojno}, {Valentini}, {Chiappini}, {Matijevi{\v{c}}}, {Siviero}, {de Laverny}, {Recio-Blanco}, {Bijaoui}, {Wyse}, {Binney}, {Grebel}, {Helmi}, {Jofre}, {Antoja}, {Gilmore}, {Siebert}, {Famaey}, {Bienaym{\'e}}, {Gibson}, {Freeman}, {Navarro}, {Munari}, {Seabroke}, {Anguiano}, {{\v{Z}}erjal}, {Minchev}, {Reid}, {Bland-Hawthorn}, {Kos}, {Sharma}, {Watson}, {Parker}, {Scholz}, {Burton}, {Cass}, {Hartley}, {Fiegert}, {Stupar}, {Ritter}, {Hawkins}, {Gerhard}, {Chaplin}, {Davies}, {Elsworth}, {Lund}, {Miglio}, \& {Mosser}}]{2017AJ....153...75K}
{Kunder}, A., {Kordopatis}, G., {Steinmetz}, M., {et~al.} 2017, \aj, 153, 75, \dodoi{10.3847/1538-3881/153/2/75}

\bibitem[{{Lai} {et~al.}(2004){Lai}, {Bolte}, {Johnson}, \& {Lucatello}}]{2004AAS...205.5210L}
{Lai}, D.~K., {Bolte}, M., {Johnson}, J.~A., \& {Lucatello}, S. 2004, in American Astronomical Society Meeting Abstracts, Vol. 205, American Astronomical Society Meeting Abstracts, 52.10

\bibitem[{{Li} {et~al.}(2018){Li}, {Tan}, \& {Zhao}}]{2018ApJS..238...16L}
{Li}, H., {Tan}, K., \& {Zhao}, G. 2018, \apjs, 238, 16, \dodoi{10.3847/1538-4365/aada4a}

\bibitem[{{Limberg} {et~al.}(2021){Limberg}, {Rossi}, {Beers}, {Perottoni}, {P{\'e}rez-Villegas}, {Santucci}, {Abuchaim}, {Placco}, {Lee}, {Christlieb}, {Norris}, {Bessell}, {Ryan}, {Wilhelm}, {Rhee}, \& {Frebel}}]{2021ApJ...907...10L}
{Limberg}, G., {Rossi}, S., {Beers}, T.~C., {et~al.} 2021, \apj, 907, 10, \dodoi{10.3847/1538-4357/abcb87}

\bibitem[{{Lu} {et~al.}(2024){Lu}, {Yuan}, {Xu}, {Zhang}, {Xiao}, {Huang}, {Beers}, \& {Hong}}]{2024ApJS..271...26L}
{Lu}, X., {Yuan}, H., {Xu}, S., {et~al.} 2024, \apjs, 271, 26, \dodoi{10.3847/1538-4365/ad1eea}

\bibitem[{{Majewski} {et~al.}(2017){Majewski}, {Schiavon}, {Frinchaboy}, {Allende Prieto}, {Barkhouser}, {Bizyaev}, {Blank}, {Brunner}, {Burton}, {Carrera}, {Chojnowski}, {Cunha}, {Epstein}, {Fitzgerald}, {Garc{\'\i}a P{\'e}rez}, {Hearty}, {Henderson}, {Holtzman}, {Johnson}, {Lam}, {Lawler}, {Maseman}, {M{\'e}sz{\'a}ros}, {Nelson}, {Nguyen}, {Nidever}, {Pinsonneault}, {Shetrone}, {Smee}, {Smith}, {Stolberg}, {Skrutskie}, {Walker}, {Wilson}, {Zasowski}, {Anders}, {Basu}, {Beland}, {Blanton}, {Bovy}, {Brownstein}, {Carlberg}, {Chaplin}, {Chiappini}, {Eisenstein}, {Elsworth}, {Feuillet}, {Fleming}, {Galbraith-Frew}, {Garc{\'\i}a}, {Garc{\'\i}a-Hern{\'a}ndez}, {Gillespie}, {Girardi}, {Gunn}, {Hasselquist}, {Hayden}, {Hekker}, {Ivans}, {Kinemuchi}, {Klaene}, {Mahadevan}, {Mathur}, {Mosser}, {Muna}, {Munn}, {Nichol}, {O'Connell}, {Parejko}, {Robin}, {Rocha-Pinto}, {Schultheis}, {Serenelli}, {Shane}, {Silva Aguirre}, {Sobeck}, {Thompson}, {Troup}, {Weinberg}, \& {Zamora}}]{2017AJ....154...94M}
{Majewski}, S.~R., {Schiavon}, R.~P., {Frinchaboy}, P.~M., {et~al.} 2017, \aj, 154, 94, \dodoi{10.3847/1538-3881/aa784d}

\bibitem[{{Martin} {et~al.}(2023){Martin}, {Starkenburg}, {Yuan}, {Fouesneau}, {Arentsen}, {De Angeli}, {Gran}, {Montelius}, {Andrae}, {Bellazzini}, {Montegriffo}, {Esselink}, {Zhang}, {Venn}, {Viswanathan}, {Aguado}, {Battaglia}, {Bayer}, {Bonifacio}, {Caffau}, {C{\^o}t{\'e}}, {Carlberg}, {Fabbro}, {Fern{\'a}ndez Alvar}, {Gonz{\'a}lez Hern{\'a}ndez}, {Gonz{\'a}lez Rivera de La Vernhe}, {Hill}, {Ibata}, {Jablonka}, {Kordopatis}, {Lardo}, {McConnachie}, {Navarrete}, {Navarro}, {Recio-Blanco}, {S{\'a}nchez Janssen}, {Sestito}, {Thomas}, {Vitali}, \& {Youakim}}]{2023arXiv230801344M}
{Martin}, N.~F., {Starkenburg}, E., {Yuan}, Z., {et~al.} 2023, arXiv e-prints, arXiv:2308.01344, \dodoi{10.48550/arXiv.2308.01344}

\bibitem[{{Mendes de Oliveira} {et~al.}(2019){Mendes de Oliveira}, {Ribeiro}, {Schoenell}, {Kanaan}, {Overzier}, {Molino}, {Sampedro}, {Coelho}, {Barbosa}, {Cortesi}, {Costa-Duarte}, {Herpich}, {Hernandez-Jimenez}, {Placco}, {Xavier}, {Abramo}, {Saito}, {Chies-Santos}, {Ederoclite}, {Lopes de Oliveira}, {Gon{\c{c}}alves}, {Akras}, {Almeida}, {Almeida-Fernandes}, {Beers}, {Bonatto}, {Bonoli}, {Cypriano}, {Vinicius-Lima}, {de Souza}, {Fabiano de Souza}, {Ferrari}, {Gon{\c{c}}alves}, {Gonzalez}, {Guti{\'e}rrez-Soto}, {Hartmann}, {Jaffe}, {Kerber}, {Lima-Dias}, {Lopes}, {Menendez-Delmestre}, {Nakazono}, {Novais}, {Ortega-Minakata}, {Pereira}, {Perottoni}, {Queiroz}, {Reis}, {Santos}, {Santos-Silva}, {Santucci}, {Barbosa}, {Siffert}, {Sodr{\'e}}, {Torres-Flores}, {Westera}, {Whitten}, {Alcaniz}, {Alonso-Garc{\'\i}a}, {Alencar}, {Alvarez-Candal}, {Amram}, {Azanha}, {Barb{\'a}}, {Bernardinelli}, {Borges Fernandes}, {Branco}, {Brito-Silva}, {Buzzo}, {Caffer}, {Campillay}, {Cano}, {Carvano}, {Castejon}, {Cid Fernandes}, {Dantas}, {Daflon}, {Damke}, {de la Reza}, {de Melo de Azevedo}, {De Paula}, {Diem}, {Donnerstein}, {Dors}, {Dupke}, {Eikenberry}, {Escudero}, {Faifer}, {Far{\'\i}as}, {Fernandes}, {Fernandes}, {Fontes}, {Galarza}, {Hirata}, {Katena}, {Gregorio-Hetem}, {Hern{\'a}ndez-Fern{\'a}ndez}, {Izzo}, {Jaque Arancibia}, {Jatenco-Pereira}, {Jim{\'e}nez-Teja}, {Kann}, {Krabbe}, {Labayru}, {Lazzaro}, {Lima Neto}, {Lopes}, {Magalh{\~a}es}, {Makler}, {de Menezes}, {Miralda-Escud{\'e}}, {Monteiro-Oliveira}, {Montero-Dorta}, {Mu{\~n}oz-Elgueta}, {Nemmen}, {Nilo Castell{\'o}n}, {Oliveira}, {Ort{\'\i}z}, {Pattaro}, {Pereira}, {Quint}, {Riguccini}, {Rocha Pinto}, {Rodrigues}, {Roig}, {Rossi}, {Saha}, {Santos}, {Schnorr M{\"u}ller}, {Sesto}, {Silva}, {Smith Castelli}, {Teixeira}, {Telles}, {Thom de Souza}, {Th{\"o}ne}, {Trevisan}, {de Ugarte Postigo}, {Urrutia-Viscarra}, {Veiga}, {Vika}, {Vitorelli}, {Werle}, {Werner}, \& {Zaritsky}}]{2019MNRAS.489..241M}
{Mendes de Oliveira}, C., {Ribeiro}, T., {Schoenell}, W., {et~al.} 2019, \mnras, 489, 241, \dodoi{10.1093/mnras/stz1985}

\bibitem[{{Nordstr{\"o}m} {et~al.}(2004){Nordstr{\"o}m}, {Andersen}, {Holmberg}, {J{\o}rgensen}, {Mayor}, \& {Pont}}]{2004PASA...21..129N}
{Nordstr{\"o}m}, B., {Andersen}, J., {Holmberg}, J., {et~al.} 2004, \pasa, 21, 129, \dodoi{10.1071/AS04013}

\bibitem[{{Rockosi} {et~al.}(2022){Rockosi}, {Lee}, {Morrison}, {Yanny}, {Johnson}, {Lucatello}, {Sobeck}, {Beers}, {Allende Prieto}, {An}, {Bizyaev}, {Blanton}, {Casagrande}, {Eisenstein}, {Gould}, {Gunn}, {Harding}, {Ivans}, {Jacobson}, {Janesh}, {Knapp}, {Kollmeier}, {L{\'e}pine}, {L{\'o}pez-Corredoira}, {Ma}, {Newberg}, {Pan}, {Prchlik}, {Sayers}, {Schlesinger}, {Simmerer}, \& {Weinberg}}]{2022ApJS..259...60R}
{Rockosi}, C.~M., {Lee}, Y.~S., {Morrison}, H.~L., {et~al.} 2022, \apjs, 259, 60, \dodoi{10.3847/1538-4365/ac5323}

\bibitem[{{Schlaufman} \& {Casey}(2014)}]{2014ApJ...797...13S}
{Schlaufman}, K.~C., \& {Casey}, A.~R. 2014, \apj, 797, 13, \dodoi{10.1088/0004-637X/797/1/13}

\bibitem[{{Schlegel} {et~al.}(1998){Schlegel}, {Finkbeiner}, \& {Davis}}]{1998ApJ...500..525S}
{Schlegel}, D.~J., {Finkbeiner}, D.~P., \& {Davis}, M. 1998, \apj, 500, 525, \dodoi{10.1086/305772}

\bibitem[{{Soubiran} {et~al.}(2016){Soubiran}, {Le Campion}, {Brouillet}, \& {Chemin}}]{2016A&A...591A.118S}
{Soubiran}, C., {Le Campion}, J.-F., {Brouillet}, N., \& {Chemin}, L. 2016, \aap, 591, A118, \dodoi{10.1051/0004-6361/201628497}

\bibitem[{{Starkenburg} {et~al.}(2017){Starkenburg}, {Martin}, {Youakim}, {Aguado}, {Allende Prieto}, {Arentsen}, {Bernard}, {Bonifacio}, {Caffau}, {Carlberg}, {C{\^o}t{\'e}}, {Fouesneau}, {Fran{\c{c}}ois}, {Franke}, {Gonz{\'a}lez Hern{\'a}ndez}, {Gwyn}, {Hill}, {Ibata}, {Jablonka}, {Longeard}, {McConnachie}, {Navarro}, {S{\'a}nchez-Janssen}, {Tolstoy}, \& {Venn}}]{2017MNRAS.471.2587S}
{Starkenburg}, E., {Martin}, N., {Youakim}, K., {et~al.} 2017, \mnras, 471, 2587, \dodoi{10.1093/mnras/stx1068}

\bibitem[{{Str{\"o}mgren}(1964)}]{1964ApNr....9..333S}
{Str{\"o}mgren}, B. 1964, Astrophysica Norvegica, 9, 333

\bibitem[{{Tan} {et~al.}(2022){Tan}, {Zhao}, {Fan}, {Wang}, {Yuan}, {Zheng}, {Li}, {Song}, \& {Zhao}}]{2022RAA....22j5004T}
{Tan}, K.-F., {Zhao}, G., {Fan}, Z., {et~al.} 2022, Research in Astronomy and Astrophysics, 22, 105004, \dodoi{10.1088/1674-4527/ac8b5b}

\bibitem[{{Wang} {et~al.}(2014){Wang}, {Zhao}, {Chen}, \& {Liu}}]{2014IAUS..298..326W}
{Wang}, W., {Zhao}, G., {Chen}, Y., \& {Liu}, Y. 2014, in Setting the scene for Gaia and LAMOST, ed. S.~{Feltzing}, G.~{Zhao}, N.~A. {Walton}, \& P.~{Whitelock}, Vol. 298, 326--330, \dodoi{10.1017/S1743921313006534}

\bibitem[{{Whitten} {et~al.}(2021){Whitten}, {Placco}, {Beers}, {An}, {Lee}, {Almeida-Fernandes}, {Herpich}, {Daflon}, {Barbosa}, {Perottoni}, {Rossi}, {Tissera}, {Yoon}, {Youakim}, {Schoenell}, {Ribeiro}, \& {Kanaan}}]{2021ApJ...912..147W}
{Whitten}, D.~D., {Placco}, V.~M., {Beers}, T.~C., {et~al.} 2021, \apj, 912, 147, \dodoi{10.3847/1538-4357/abee7e}

\bibitem[{{Xu} {et~al.}(2022){Xu}, {Yuan}, {Zhang}, {Li}, {Beers}, \& {Huang}}]{2022ApJS..263...29X}
{Xu}, S., {Yuan}, H., {Zhang}, R., {et~al.} 2022, \apjs, 263, 29, \dodoi{10.3847/1538-4365/ac9908}

\bibitem[{{Yang} {et~al.}(2022){Yang}, {Yuan}, {Xiang}, {Duan}, {Huang}, {Liu}, {Beers}, {Galarza}, {Daflon}, {Fern{\'a}ndez-Ontiveros}, {Cenarro}, {Crist{\'o}bal-Hornillos}, {Hern{\'a}ndez-Monteagudo}, {L{\'o}pez-Sanjuan}, {Mar{\'\i}n-Franch}, {Moles}, {Varela}, {V{\'a}zquez Rami{\'o}}, {Alcaniz}, {Dupke}, {Ederoclite}, {Sodr{\'e}}, \& {Angulo}}]{2022A&A...659A.181Y}
{Yang}, L., {Yuan}, H., {Xiang}, M., {et~al.} 2022, \aap, 659, A181, \dodoi{10.1051/0004-6361/202142724}

\bibitem[{{Yanny} {et~al.}(2009){Yanny}, {Rockosi}, {Newberg}, {Knapp}, {Adelman-McCarthy}, {Alcorn}, {Allam}, {Allende Prieto}, {An}, {Anderson}, {Anderson}, {Bailer-Jones}, {Bastian}, {Beers}, {Bell}, {Belokurov}, {Bizyaev}, {Blythe}, {Bochanski}, {Boroski}, {Brinchmann}, {Brinkmann}, {Brewington}, {Carey}, {Cudworth}, {Evans}, {Evans}, {Gates}, {G{\"a}nsicke}, {Gillespie}, {Gilmore}, {Nebot Gomez-Moran}, {Grebel}, {Greenwell}, {Gunn}, {Jordan}, {Jordan}, {Harding}, {Harris}, {Hendry}, {Holder}, {Ivans}, {Ivezi{\v{c}}}, {Jester}, {Johnson}, {Kent}, {Kleinman}, {Kniazev}, {Krzesinski}, {Kron}, {Kuropatkin}, {Lebedeva}, {Lee}, {French Leger}, {L{\'e}pine}, {Levine}, {Lin}, {Long}, {Loomis}, {Lupton}, {Malanushenko}, {Malanushenko}, {Margon}, {Martinez-Delgado}, {McGehee}, {Monet}, {Morrison}, {Munn}, {Neilsen}, {Nitta}, {Norris}, {Oravetz}, {Owen}, {Padmanabhan}, {Pan}, {Peterson}, {Pier}, {Platson}, {Re Fiorentin}, {Richards}, {Rix}, {Schlegel}, {Schneider}, {Schreiber}, {Schwope}, {Sibley}, {Simmons}, {Snedden}, {Allyn Smith}, {Stark}, {Stauffer}, {Steinmetz}, {Stoughton}, {SubbaRao}, {Szalay}, {Szkody}, {Thakar}, {Sivarani}, {Tucker}, {Uomoto}, {Vanden Berk}, {Vidrih}, {Wadadekar}, {Watters}, {Wilhelm}, {Wyse}, {Yarger}, \& {Zucker}}]{2009AJ....137.4377Y}
{Yanny}, B., {Rockosi}, C., {Newberg}, H.~J., {et~al.} 2009, \aj, 137, 4377, \dodoi{10.1088/0004-6256/137/5/4377}

\bibitem[{{Yong} {et~al.}(2013){Yong}, {Norris}, {Bessell}, {Christlieb}, {Asplund}, {Beers}, {Barklem}, {Frebel}, \& {Ryan}}]{2013ApJ...762...26Y}
{Yong}, D., {Norris}, J.~E., {Bessell}, M.~S., {et~al.} 2013, \apj, 762, 26, \dodoi{10.1088/0004-637X/762/1/26}

\bibitem[{{Yoon} {et~al.}(2016){Yoon}, {Beers}, {Placco}, {Rasmussen}, {Carollo}, {He}, {Hansen}, {Roederer}, \& {Zeanah}}]{2016ApJ...833...20Y}
{Yoon}, J., {Beers}, T.~C., {Placco}, V.~M., {et~al.} 2016, \apj, 833, 20, \dodoi{10.3847/0004-637X/833/1/20}

\bibitem[{{York} {et~al.}(2000){York}, {Adelman}, {Anderson}, {Anderson}, {Annis}, {Bahcall}, {Bakken}, {Barkhouser}, {Bastian}, {Berman}, {Boroski}, {Bracker}, {Briegel}, {Briggs}, {Brinkmann}, {Brunner}, {Burles}, {Carey}, {Carr}, {Castander}, {Chen}, {Colestock}, {Connolly}, {Crocker}, {Csabai}, {Czarapata}, {Davis}, {Doi}, {Dombeck}, {Eisenstein}, {Ellman}, {Elms}, {Evans}, {Fan}, {Federwitz}, {Fiscelli}, {Friedman}, {Frieman}, {Fukugita}, {Gillespie}, {Gunn}, {Gurbani}, {de Haas}, {Haldeman}, {Harris}, {Hayes}, {Heckman}, {Hennessy}, {Hindsley}, {Holm}, {Holmgren}, {Huang}, {Hull}, {Husby}, {Ichikawa}, {Ichikawa}, {Ivezi{\'c}}, {Kent}, {Kim}, {Kinney}, {Klaene}, {Kleinman}, {Kleinman}, {Knapp}, {Korienek}, {Kron}, {Kunszt}, {Lamb}, {Lee}, {Leger}, {Limmongkol}, {Lindenmeyer}, {Long}, {Loomis}, {Loveday}, {Lucinio}, {Lupton}, {MacKinnon}, {Mannery}, {Mantsch}, {Margon}, {McGehee}, {McKay}, {Meiksin}, {Merelli}, {Monet}, {Munn}, {Narayanan}, {Nash}, {Neilsen}, {Neswold}, {Newberg}, {Nichol}, {Nicinski}, {Nonino}, {Okada}, {Okamura}, {Ostriker}, {Owen}, {Pauls}, {Peoples}, {Peterson}, {Petravick}, {Pier}, {Pope}, {Pordes}, {Prosapio}, {Rechenmacher}, {Quinn}, {Richards}, {Richmond}, {Rivetta}, {Rockosi}, {Ruthmansdorfer}, {Sandford}, {Schlegel}, {Schneider}, {Sekiguchi}, {Sergey}, {Shimasaku}, {Siegmund}, {Smee}, {Smith}, {Snedden}, {Stone}, {Stoughton}, {Strauss}, {Stubbs}, {SubbaRao}, {Szalay}, {Szapudi}, {Szokoly}, {Thakar}, {Tremonti}, {Tucker}, {Uomoto}, {Vanden Berk}, {Vogeley}, {Waddell}, {Wang}, {Watanabe}, {Weinberg}, {Yanny}, {Yasuda}, \& {SDSS Collaboration}}]{2000AJ....120.1579Y}
{York}, D.~G., {Adelman}, J., {Anderson}, John~E., J., {et~al.} 2000, \aj, 120, 1579, \dodoi{10.1086/301513}

\bibitem[{{Yuan} {et~al.}(2015){Yuan}, {Liu}, {Xiang}, {Huang}, \& {Chen}}]{2015ApJ...803...13Y}
{Yuan}, H., {Liu}, X., {Xiang}, M., {Huang}, Y., \& {Chen}, B. 2015, \apj, 803, 13, \dodoi{10.1088/0004-637X/803/1/13}

\bibitem[{{Zhang} \& {Yuan}(2023)}]{2023ApJS..264...14Z}
{Zhang}, R., \& {Yuan}, H. 2023, \apjs, 264, 14, \dodoi{10.3847/1538-4365/ac9dfa}

\bibitem[{{Zhang} {et~al.}(2021){Zhang}, {Yuan}, {Liu}, {Xiang}, {Huang}, \& {Chen}}]{2021RAA....21..319Z}
{Zhang}, R.-Y., {Yuan}, H.-B., {Liu}, X.-W., {et~al.} 2021, Research in Astronomy and Astrophysics, 21, 319, \dodoi{10.1088/1674-4527/21/12/319}

\bibitem[{{Zhao} {et~al.}(2012){Zhao}, {Zhao}, {Chu}, {Jing}, \& {Deng}}]{2012RAA....12..723Z}
{Zhao}, G., {Zhao}, Y.-H., {Chu}, Y.-Q., {Jing}, Y.-P., \& {Deng}, L.-C. 2012, Research in Astronomy and Astrophysics, 12, 723, \dodoi{10.1088/1674-4527/12/7/002}

\bibitem[{{Zheng} {et~al.}(2018){Zheng}, {Zhao}, {Wang}, {Fan}, {Tan}, {Li}, \& {Zuo}}]{2018RAA....18..147Z}
{Zheng}, J., {Zhao}, G., {Wang}, W., {et~al.} 2018, Research in Astronomy and Astrophysics, 18, 147, \dodoi{10.1088/1674-4527/18/12/147}

\bibitem[{{Zheng} {et~al.}(2019){Zheng}, {Zhao}, {Wang}, {Fan}, {Tan}, {Li}, \& {Zuo}}]{2019RAA....19....3Z}
---. 2019, Research in Astronomy and Astrophysics, 19, 003, \dodoi{10.1088/1674-4527/19/1/3}

\end{thebibliography}
\bibliographystyle{aasjournal}

\end{document}